\documentclass{iopjournal}
\usepackage{cmap}
\usepackage[utf8]{inputenc}
\usepackage[T1]{fontenc}
\usepackage{lmodern}
\usepackage{amsmath,amssymb}
\usepackage[numbers,sort&compress]{natbib}

\def\imo{i}
\def\re#1{\mathrm{Re}\left(#1\right)}
\def\im#1{\mathrm{Im}\left(#1\right)}
\def\K{\mathcal{K}}
\def\V{\widetilde{V}}
\def\Order#1{\mathcal{O}\left(#1\right)}
\def\order#1{{o}\left(#1\right)}

\begin{document}
\title{Grey-body factors of black holes}
\author{Roman A. Konoplya$^{1,2}$\orcid{0000-0003-1343-9584} and Alexander Zhidenko\orcid{0000-0001-6838-3309}$^{1,3}$}

\affil{$^1$ Research Centre for Theoretical Physics and Astrophysics, Institute of Physics, Silesian University in Opava,\\ Bezručovo náměstí 1150/13, CZ-74601 Opava, Czech Republic}

\affil{$^2$ Department of Physics, Faculty of Science, University of Hradec Králové,\\ Rokitanského 62/26, 500 03 Hradec Králové, Czech Republic}

\affil{$^3$ Centro de Matemática, Computação e Cognição (CMCC), Universidade Federal do ABC (UFABC),\\ Rua Abolição, CEP: 09210-180, Santo André, SP, Brazil}


\begin{abstract}
Grey-body factors describe the frequency-dependent transmission of fields through the effective potential generated by a black hole or another compact object. They convert the idealized thermal spectrum of Hawking radiation into the spectrum measured by an observer, determine absorption and scattering cross sections, and provide real-frequency observables complementary to quasinormal modes. In this review, we present a unified account of grey-body factors in black-hole physics and related systems. We formulate the scattering problem for asymptotically flat, de Sitter, and anti-de Sitter spacetimes, and discuss the connections between transmission coefficients, absorption cross sections, Hawking radiation, and ringdown. We also review methods for their computation and survey recent developments, including the correspondence between grey-body factors and quasinormal modes, their stability under deformations of the effective potential, wormhole transmission and superradiance, and their behavior in modified theories of gravity. Finally, we discuss the appearance of grey-body factors in gauge/gravity duality, where absorption probabilities in the bulk are related to correlation functions and transport coefficients in the boundary theory.
\end{abstract}

\bigskip
\centerline{This review is dedicated to the memory of Professor Jan Kříž (1974-2026), our colleague and friend.}

\tableofcontents
\par\noindent\rule{\linewidth}{0.4pt}

\section{Introduction}\label{sec:introduction}
In gravitational physics, grey-body factors arise whenever radiation emitted near a compact object must propagate through the effective potential created by spacetime curvature. In black-hole thermodynamics they correct the idealized Hawking spectrum, determining the true particle fluxes and evaporation rates. They also enter the description of absorption and scattering cross sections, where the black hole acts as a frequency-dependent filter for incoming waves, and they are linked to quasinormal modes through resonances of the scattering matrix. Similar effects occur for neutron stars and other compact objects, where the external geometry and matter distribution imprint frequency-dependent signatures on the emitted or absorbed radiation.

The role of grey-body factors, however, extends far beyond gravity. In astrophysics, the observed spectra of stars are shaped by the transmission properties of their atmospheres, which act as natural grey-body filters. In quantum optics and condensed matter systems, cavities, photonic crystals, and metamaterials distort otherwise thermal emission in precisely the same way. Even in nuclear and particle physics, thermal-like radiation from quark--gluon plasma or hot nuclear matter is filtered by the surrounding medium, leading to effective grey-body corrections. Thus, grey-body factors provide a unifying language for describing how any thermal source, whether gravitational, astrophysical, or laboratory, is spectrally reshaped by the medium or potential between emission and observation.

Historically, the concept underlying grey-body factors predates black-hole physics and can be traced back to the theory of thermal radiation and wave scattering. In classical thermodynamics, it was already recognized that realistic bodies do not emit as perfect blackbodies, but rather with a frequency-dependent emissivity \cite{Kirchhoff1860}. This leads to a modification of the Planck spectrum by a factor $\Gamma(\omega)$,
\begin{equation}
I(\omega)=\Gamma(\omega)\,I_{\rm Planck}(\omega),
\end{equation}
where $\Gamma(\omega)\leq 1$ accounts for absorption and reflection in the material. Although introduced in a different physical context, this frequency-dependent correction anticipates the modern notion of grey-body factors as transmission probabilities through an effective barrier.

A more direct precursor arises in quantum mechanics, where scattering by a potential barrier is characterized by reflection and transmission coefficients. In particular, the tunneling probability that Gamow derived in \cite{Gamow:1928} for $\alpha$-decay provides a prototypical example of a frequency-dependent transmission coefficient,
\begin{equation}
\Gamma(\omega)\equiv |T(\omega)|^2,
\end{equation}
determined by the structure of the potential barrier. In the language of partial waves, each angular mode $\ell$ (including the $(2 \ell+1)$-fold degeneracy of the azimuthal number $m$) contributes independently, and the total transmission is governed by the effective potential felt by the wave. This formalism, developed for nuclear and wave scattering problems, is directly applicable to fields propagating in curved spacetimes.

The notion of grey-body factors in black-hole physics emerged when Hawking radiation was supplemented by the effect of spacetime curvature outside the horizon. While the radiation is thermal at the horizon, it is filtered by the effective potential barrier surrounding the black hole. The first detailed calculations of these transmission coefficients were performed by Page \cite{Page:1976df}, who showed how the spectrum deviates from a pure blackbody. At the same time, studies of wave scattering by black holes demonstrated that the absorption cross section is governed by the same transmission probabilities \cite{Sanchez:1976,Sanchez:1978}.
Grey-body factors also played an important historical role in the emergence of gauge/gravity duality. The agreement between the grey-body spectra of near-extremal black holes and the corresponding D-brane absorption probabilities \cite{Maldacena:1996ix} was among the first concrete indications that gravitational radiation processes have a dual quantum/string-theory description, paving the way toward the formulation of the AdS/CFT correspondence.

It is useful to stress from the outset that, in black-hole physics, a grey-body factor is not a phenomenological emissivity inserted by hand. It is a mode-dependent scattering probability fixed by the field equation, the boundary conditions at the relevant horizons or asymptotic regions, and the quantum numbers of the perturbing field. This mode-by-mode character explains why accurate grey-body factors are indispensable in phenomenological applications. In evaporation calculations, the observable spectrum depends on the spin, mass threshold, statistics, internal degeneracy, and species-dependent transmission probability of each emitted particle. Approximating the transmission probability by unity or by a sharp geometric-optics cutoff can distort both the total lifetime and the observable spectra. At the same time, because grey-body factors are defined on the real-frequency axis, they provide scattering observables that can complement quasinormal modes and enter frequency-domain descriptions of the ringdown \cite{Oshita:2023cjz,Okabayashi:2024qbz,Rosato:2026apq}.

Our review focuses on a more specific subject: grey-body factors of black holes and other compact objects for which analogous definitions can be formulated, such as black branes and wormholes. Grey-body factors are essential for estimating the intensity of Hawking radiation and can, in principle, also be applied to a broad class of radiative processes occurring in the vicinity of compact objects.

The review is organized as follows. In Sec.~\ref{sec:scattering} we formulate the scattering problem and introduce the definition of grey-body factors under asymptotically flat, de Sitter, and anti-de Sitter boundary conditions. Section~\ref{sec:applications} discusses physical applications of grey-body factors, including absorption cross sections, Hawking radiation, and black-hole ringdown. In Sec.~\ref{sec:analytic} we review the principal analytic and semi-analytic techniques for calculating grey-body factors, including the WKB approximation, eikonal, low- and high-frequency asymptotics, Frobenius and inverse scattering methods, while Sec.~\ref{sec:Visser} is devoted to the Visser bound. Numerical approaches, publicly available codes, and spectral methods are discussed in Sec.~\ref{sec:numerical}. Section~\ref{sec:correspondence} reviews the recently established correspondence between grey-body factors and quasinormal modes. In Sec.~\ref{sec:blackholes} we survey grey-body factors for a broad class of compact objects and gravitational theories, including wormholes and modified-gravity black holes, while Sec.~\ref{sec:duality} discusses their appearance in the context of gauge/gravity duality. Finally, Sec.~\ref{sec:summary} summarizes the main results and outlines open problems and future directions.
Throughout this review, unless dimensional units are restored explicitly, we use natural units $G=c=\hbar=k_{\rm B}=1$.

\section{The scattering problem}\label{sec:scattering}

In this section we formulate the scattering setup used throughout the review and define the corresponding grey-body factors under different asymptotic boundary conditions.

\subsection{Master equation}

For stationary axisymmetric backgrounds, the perturbation equation is always Fourier-decomposable with respect to $t$ and $\phi$, but is not, in general, separable in $(r,\theta)$. A generic partial decomposition can be written as
\begin{equation}
\varphi(t,r,\theta,\phi)=e^{-i\omega t}e^{im\phi}\,\psi_{\omega m}(r,\theta),
\end{equation}
where $\omega\in\mathbb{R}$ is the scattering frequency and $m$ is the azimuthal number. In the general case, $\psi_{\omega m}(r,\theta)$ represents several coupled perturbation amplitudes. When possible, the corresponding field equations are decoupled by forming gauge-invariant combinations of these amplitudes, each of which satisfies a separate master equation.

The mode ansatz itself does not impose any sign restriction on $\omega$. In what follows we impose the convention $\omega>0$ for the scattering problem. This choice follows from the particle definition at infinity: positive Killing frequency with respect to $t$ labels physical outgoing quanta. Modes with $\omega<0$ are not independent degrees of freedom, because they are related to positive-frequency modes by complex/charge conjugation with opposite quantum numbers, i.e. $m\to -m$ and, for charged fields, $q\to -q$. Therefore, including both signs would double count states.

Separation in $r$ and $\theta$ becomes possible only for special classes of backgrounds and perturbation equations. For spherically symmetric backgrounds, separation follows from the expansion in spherical or hyperspherical harmonics. For the rotating backgrounds, it is possible only in special cases, typically when a hidden symmetry (such as a Carter constant, a Killing tensor, or an underlying Killing--Yano structure) is inherited by the field equation \cite{Carter:1968rr,Carter:1968ks,Walker:1970un,Frolov:2006dqt,Chandrasekhar:1983,Konoplya:2018arm}. Even in such a geometry, separability must be established independently for each field spin and coupling: separability of the Hamilton--Jacobi or Klein--Gordon equation does not, by itself, imply separability of electromagnetic, gravitational, or coupled perturbations. Whenever the required separability holds, one may set
\begin{equation}\label{angular}
\psi_{\omega m}(r,\theta)=S_{\ell m}(\theta)R_{\ell m}(r),
\end{equation}
where $\ell$ is the multipole number.

Examples include scalar fields in Kerr and Kerr--Newman geometries and electromagnetic and gravitational perturbations of Kerr governed by the Teukolsky equations \cite{Teukolsky:1972my,Teukolsky:1973ha}, as well as the corresponding nonrotating limits. The ansatz (\ref{angular}) is not generally applicable to coupled gravitoelectromagnetic perturbations of Kerr--Newman, to systems involving additional dynamical fields unless a further decoupling is possible, or to generic rotating deformations for which different angular multipoles remain coupled. In these nonseparable cases, one must instead solve a system of equations in $(r,\theta)$ or expand the angular dependence in a suitable basis and solve the resulting coupled radial equations. In the separable case, substitution into the master equation yields an angular spectral problem,
\begin{equation}
\mathcal{L}_{\theta}\,S_{\ell m}=-\Lambda_{\ell m}S_{\ell m},
\end{equation}
where $\Lambda_{\ell m}$ is the separation constant, and a radial equation in Schrödinger-like form,
\begin{equation}\label{schroedinger_GBF}
\frac{d^2\Psi}{dr_*^2}+U(\Lambda_{\ell m},\omega;r_*)\Psi=0,
\end{equation}
where $\Psi\propto R_{\ell m}(r)$ and $r_*$ is the tortoise coordinate, defined so that the $(t,r_*)$ sector becomes conformally flat. For separable backgrounds, $r_*$ depends only on $r$ and can be introduced as
\begin{equation}\label{tortoise}
dr_*=\frac{dr}{F(r)},
\end{equation}
where $F(r)$ depends on the specific geometry.

One has $r_*\to-\infty$ at the event horizon. In asymptotically flat spacetimes, $r_*\to\infty$ as $r\to\infty$. For de Sitter black holes, $r_*\to +\infty$ at the cosmological horizon, whereas in asymptotically AdS cases $r_*$ approaches a finite limit at the AdS boundary.

\subsection{Scattering problem and grey-body factor}
The function $U$ approaches constants near both asymptotic regions. At the event horizon,
\begin{equation}
\lim_{r_*\to-\infty}U=(\omega-\Phi_H)^2,
\end{equation}
where $\Phi_H$ depends on the particular configuration: for a rotating black hole, $\Phi_H\equiv m\Omega_H$, where $\Omega_H$ is the horizon angular velocity; for a charged black hole, $\Phi_H$ is the Coulomb energy of the field particle at the horizon.

For an asymptotically flat spacetime,
\begin{equation}
\lim_{r_*\to\infty}U=\omega^2-\mu^2\equiv k^2,
\end{equation}
satisfying the usual dispersion relation, where $\mu$ is the mass of the field under consideration. Conventionally, we take $k>0$, so that for massless perturbations ($\mu=0$) one has $\omega=k$.

The scattering problem is then well posed by imposing unit-amplitude incidence from spatial infinity together with a purely ingoing condition at the event horizon (classically, no outgoing flux can emerge from the horizon):
\begin{equation}\label{eq:scattering}
\begin{array}{rclcl}
\Psi &\to& e^{-ik r_*}+R_{\ell m}e^{+ik r_*}, &\quad& r_*\to+\infty, \\
\Psi &\to& T_{\ell m}e^{-i(\omega-\Phi_H)r_*}, &\quad& r_*\to-\infty,
\end{array}
\end{equation}
where the incident wave has unit amplitude, $R_{\ell m}$ is the reflection coefficient, and $T_{\ell m}$ is the transmission coefficient.

When Eq.~(\ref{schroedinger_GBF}) has real coefficients, the complex-conjugate solution $\Psi^*$ is also a solution, and the corresponding Wronskian
\begin{equation}\label{Wronskian}
W(r_*) \equiv \Psi\frac{d\Psi^*}{dr_*}-\Psi^*\frac{d\Psi}{dr_*},
\end{equation}
is independent of $r_*$ because Eq.~(\ref{schroedinger_GBF}) implies $dW/dr_*=0$. Evaluating this constant at $r_*\to+\infty$ and $r_*\to-\infty$ with the asymptotic forms above yields flux conservation and, therefore,
\begin{equation}
1-|R_{\ell m}|^2=\frac{\omega-\Phi_H}{k}\,|T_{\ell m}|^2.
\end{equation}

For a given $\omega$, the grey-body factor is defined as the transmitted-energy fraction \cite{Nakamura:1976nc}:
\begin{equation}\label{eq:greybody}
\Gamma_{\ell m}(\omega)\equiv1-|R_{\ell m}|^2=\frac{\omega-\Phi_H}{k}\,|T_{\ell m}|^2.
\end{equation}

These relations immediately imply that in the regime $0<\omega<\Phi_H$ the horizon flux is negative, while the reflected flux is larger than the incident one ($|R_{\ell m}|^2>1$). This amplification is the superradiant effect: the scattered wave extracts energy from the black hole and returns with increased amplitude \cite{Zeldovich:1971,Zeldovich:1972,Starobinsky:1973a,Starobinsky:1973b,Bekenstein:1973mi}.

For rotating black holes, where $\Phi_H=m\Omega_H$, superradiance occurs for co-rotating modes ($m>0$), i.e.~for waves carrying angular momentum in the same direction as the hole. For charged fields in charged black-hole backgrounds, an analogous mechanism extracts Coulomb energy when the field and black-hole charges have the same sign \cite{Bekenstein:1973mi,Nakamura:1976nc}.

Classical superradiant amplification is absent for fermionic (Dirac) fields: the structure of the Dirac current implies no reflected-flux enhancement of the form $|R_{\ell m}|^2>1$. Accordingly, the Wronskian argument above, based on a real Schrödinger-type equation, is not directly applicable in the fermionic case, where the effective radial equation is generally complex and $\Psi^*$ is not an independent solution.

\subsection{Reciprocal scattering problem}

Within the same real-equation setting, one can formulate the reciprocal problem with incidence from the near-horizon side, satisfying the boundary conditions~\cite{Harmark:2007jy}
\begin{equation}
\begin{array}{rclcl}
\Psi^{\prime} &\to& e^{+i(\omega-\Phi_H)r_*}+R^{\prime}_{\ell m}e^{-i(\omega-\Phi_H)r_*}, &\quad& r_*\to-\infty,\\
\Psi^{\prime} &\to& T^{\prime}_{\ell m}e^{+ik r_*}, &\quad& r_*\to+\infty.
\end{array}
\end{equation}
Comparing the Wronskian at both asymptotics, one obtains
\begin{equation}
\Gamma_{\ell m}^{\prime}(\omega)\equiv1-|R_{\ell m}^{\prime}|^2=\frac{k}{\omega-\Phi_H}\,|T_{\ell m}^{\prime}|^2.
\end{equation}

Because $\Psi$ and $\Psi^*$ form a basis of solutions to Eq.~(\ref{schroedinger_GBF}), the solution $\Psi^{\prime}$ can be written as their linear combination. Taking the limit $r_*\to-\infty$, one finds
\begin{equation}
\Psi^{\prime}=\frac{1}{T_{\ell m}^*}\Psi^*+\frac{R^{\prime}_{\ell m}}{T_{\ell m}}\Psi.
\end{equation}
Then taking the limit $r_*\to\infty$ on both sides yields
\begin{eqnarray}
\label{transmissioncompare}
T^{\prime}_{\ell m}&=&\frac{1}{T_{\ell m}^*}+\frac{R^{\prime}_{\ell m}R_{\ell m}}{T_{\ell m}},\\
\label{reflectioncompare}
0&=&\frac{R_{\ell m}^*}{T_{\ell m}^*}+\frac{R^{\prime}_{\ell m}}{T_{\ell m}}.
\end{eqnarray}
From Eq.~(\ref{reflectioncompare}) it follows that
\begin{equation}
|R^{\prime}_{\ell m}|^2=|R_{\ell m}|^2.
\end{equation}
Substituting Eq.~(\ref{reflectioncompare}) into Eq.~(\ref{transmissioncompare}), one gets
\begin{equation}
T^{\prime}_{\ell m}=\frac{1-|R_{\ell m}|^2}{T_{\ell m}^*}=\frac{\omega-\Phi_H}{k}T_{\ell m}.
\end{equation}

Thus, the grey-body factors for both problems are identical:
\begin{equation}\label{gbidentity}
\Gamma_{\ell m}^{\prime}(\omega)=\frac{k}{\omega-\Phi_H}\,|T_{\ell m}^{\prime}|^2=\frac{\omega-\Phi_H}{k}|T_{\ell m}|^2=\Gamma_{\ell m}(\omega).
\end{equation}

This identity is directly relevant for Hawking-radiation calculations. Hawking quanta are produced in the near-horizon region, whereas the observable flux is measured at infinity; the equality~(\ref{gbidentity}) guarantees that one may compute the transmission factor in either formulation and insert the same quantity into the emission spectrum. In practice, one usually determines $\Gamma_{\ell m}(\omega)$ from the standard scattering problem with incidence from infinity and then uses it as the grey-body filter in the Hawking flux formulas. This avoids solving a separate emission-side boundary-value problem and makes detailed-balance arguments transparent, including in superradiant channels where the sign of $\omega-\Phi_H$ is essential.

\subsection{Asymptotically de Sitter spacetime}

For asymptotically de Sitter spacetime, the asymptotic region $r_*\to\infty$ corresponds to the cosmological horizon rather than spatial infinity. Therefore,
\begin{equation}
\lim_{r_*\to\infty}U=(\omega-\Phi_C)^2.
\end{equation}
As in the event-horizon case, $\Phi_C$ denotes a constant shift determined by the background and the field, now evaluated at the cosmological horizon. For a rotating spacetime, one may have $\Phi_C=m\Omega_C$, where $\Omega_C$ is the angular velocity of the cosmological horizon. For a charged field in a charged background, $\Phi_C$ is the Coulomb energy of the field particle at the cosmological horizon. Thus, $\Phi_C$ plays the same role at the cosmological horizon as $\Phi_H$ does at the event horizon. The boundary conditions then read
\begin{equation}
\begin{array}{rclcl}
\Psi &\to& e^{-i(\omega-\Phi_C) r_*}+R_{\ell m}e^{+i(\omega-\Phi_C) r_*}, &\quad& r_*\to+\infty, \\
\Psi &\to& T_{\ell m}e^{-i(\omega-\Phi_H)r_*}, &\quad& r_*\to-\infty.
\end{array}
\end{equation}
Similarly, for the grey-body factor,
\begin{equation}\label{eq:greybodydS}
\Gamma_{\ell m}(\omega)\equiv1-|R_{\ell m}|^2=\frac{\omega-\Phi_H}{\omega-\Phi_C}\,|T_{\ell m}|^2.
\end{equation}
Hence, superradiance occurs in the interval
$$
\min(\Phi_H,\Phi_C)<\omega<\max(\Phi_H,\Phi_C).
$$

\subsection{Grey-body factor for given frequency}

Equation~(\ref{schroedinger_GBF}) is linear, so its overall normalization is arbitrary. A convenient parametrization of the asymptotic amplitudes is
\begin{equation}
\begin{array}{rcl}
\Psi(r_*\to+\infty) &=& A_{in}\Psi_{in}(r_*)+A_{out}\Psi_{out}(r_*), \\
\Psi(r_*\to-\infty) &=& A_{tran}\Psi_{tran}(r_*),
\end{array}
\end{equation}
where the ingoing, outgoing, and transmitted waves are
\begin{eqnarray}
\Psi_{in}(r_*)&=&\left\{
\begin{array}{ll}
e^{-ik r_*} & \mbox{(flat)},\\
e^{-i(\omega-\Phi_C) r_*} & \mbox{(de Sitter)},\\
\end{array}\right.\\
\Psi_{out}(r_*)&=&\left\{
\begin{array}{ll}
e^{+ik r_*} & \mbox{(flat)},\\
e^{+i(\omega-\Phi_C) r_*} & \mbox{(de Sitter)},\\
\end{array}\right.\\
\Psi_{tran}(r_*)&=&e^{-i(\omega-\Phi_H)r_*}.
\end{eqnarray}

Because $\Psi_{tran}(r_*)$ is well defined near the event horizon and can be computed using a series expansion, the problem is practically reduced to an initial-value problem posed close to the horizon. For a fixed frequency $\omega$, one may set $A_{tran}=1$, integrate outward, and determine $A_{in}$ and $A_{out}$ by matching the numerical solution to its asymptotic form at large $r_*$. Dividing by $A_{in}$ restores the standard unit-incident-wave normalization, and therefore
\begin{equation}\label{GBcalc}
R_{\ell m}=\frac{A_{out}}{A_{in}} \qquad\Longrightarrow\qquad \Gamma_{\ell m}(\omega)=1-\left|\frac{A_{out}}{A_{in}}\right|^2.
\end{equation}

\subsection{Asymptotically anti-de Sitter spacetime}
In the AdS spacetime,
\begin{equation}\label{FinAdS}
F(r)=r^2\left(\frac{1}{{L^2}}+\Order{\frac{1}{r}}\right),
\end{equation}
where $L$ is the AdS radius. Equation~(\ref{tortoise}) implies that in the limit $r\to\infty$ the tortoise coordinate approaches a finite value, corresponding to the AdS boundary. This boundary is a regular singular point of Eq.~(\ref{schroedinger_GBF}). The associated indicial equation has, in general, two real roots, so the asymptotic solution behaves as
\begin{equation}\label{asymptoticbehavior}
\Psi \sim C_1 r^{r_1} + C_2 r^{r_2},\qquad r_1>r_2, \quad r_1+r_2=-1.
\end{equation}
Here we have discarded terms that are subdominant at large $r$.

Equation~(\ref{asymptoticbehavior}) cannot be directly interpreted as a superposition of oscillatory ingoing and outgoing waves. Nevertheless, as suggested in \cite{Harmark:2007jy}, one can introduce two asymptotic solutions, $\Psi_{in}$ and $\Psi_{out}$, associated with purely ingoing and purely outgoing flux, respectively.

The key observation is that in asymptotically flat and de Sitter cases the Wronskian~(\ref{Wronskian}) is proportional to the momentum flux and, for pure plane waves,
\begin{equation}
\begin{array}{rclr}
W(r_*) &=& +2ik & \mbox{(outgoing)}, \\
W(r_*) &=& -2ik & \mbox{(ingoing)},
\end{array}
\end{equation}
one can impose the following conditions for the solutions $\Psi_{in}$ and $\Psi_{out}$:
\begin{equation}\label{AdSinout}
\frac{1}{2i}\left(\Psi_{out}\frac{d\Psi_{out}^*}{dr_*}-\Psi_{out}^*\frac{d\Psi_{out}}{dr_*}\right)=-\frac{1}{2i}\left(\Psi_{in}\frac{d\Psi_{in}^*}{dr_*}-\Psi_{in}^*\frac{d\Psi_{in}}{dr_*}\right)\propto k>0.
\end{equation}

The functions $\Psi_{in}$ and $\Psi_{out}$, satisfying Eqs.~(\ref{asymptoticbehavior}) and (\ref{AdSinout}), can be chosen as
\begin{eqnarray}
\Psi_{in}&\sim&A\left(r^{r_1}-iK r^{r_2}\right),\\
\Psi_{out}&\sim&A\left(r^{r_1}+iK r^{r_2}\right),
\end{eqnarray}
where $A$ is a complex normalization constant and $K>0$.
Matching the superposition of ingoing and outgoing solutions to the asymptotic behavior~(\ref{asymptoticbehavior}),
\begin{equation}
\Psi = A_{in}\Psi_{in}+A_{out}\Psi_{out} \sim C_1 r^{r_1} + C_2 r^{r_2},
\end{equation}
one finds that
$$C_1=A(A_{in}+A_{out}), \qquad C_2=-i K A(A_{in}-A_{out}).$$
The grey-body factor is then defined similarly to Eq.~(\ref{GBcalc}):
\begin{equation}\label{GBAdS}
\Gamma_{\ell m}(\omega)=1-\left|\frac{A_{out}}{A_{in}}\right|^2=1-\left|\frac{1-z(\omega)}{1+z(\omega)}\right|^2, \qquad z(\omega)=-\frac{iC_1K}{C_2}.
\end{equation}

By analyzing particular field equations for spherically symmetric AdS black holes in various dimensions, \cite{Harmark:2007jy} showed that the ingoing and outgoing fluxes in the asymptotic region can be expressed through Hankel functions,
\begin{eqnarray}
\Psi_{out}(r)&=&\left(\frac{\omega L^2}{r}\right)^{-r_1}H^{(1)}_{(r_1-r_2)/2}\left(\frac{\omega L^2}{r}\right),\\
\Psi_{in}(r)&=&\left(\frac{\omega L^2}{r}\right)^{-r_1}H^{(2)}_{(r_1-r_2)/2}\left(\frac{\omega L^2}{r}\right),
\end{eqnarray}
and therefore,
\begin{equation}
K=\frac{\pi\omega^{r_1-r_2}L^{2(r_1-r_2)}}{\Gamma\!\left(\dfrac{r_1-r_2}{2}\right)\Gamma\!\left(\dfrac{r_1-r_2}{2}+1\right)2^{r_1-r_2}}.
\end{equation}
It was also shown that $z(\omega)$ is real and positive, so the grey-body factor~(\ref{GBAdS}) obeys
\begin{equation}
0\leq \Gamma_{\ell m}(\omega) < 1.
\end{equation}

Unlike at an asymptotically flat end, the two power-law solutions at the AdS boundary do not by themselves select a unique ingoing/outgoing basis. The constant $K$ specifies the relative normalization of the two falloffs and therefore forms part of the scattering prescription. Ref.~\cite{Harmark:2007jy} removes this freedom for the class of black holes considered there by matching the asymptotic solution to a definite pair of Hankel functions. A similar analysis for the rotating linear dilaton black hole was carried out in \cite{Li:2012zj}.

\subsection{Spherically symmetric configurations}
The above derivations and definitions are valid for generic stationary configurations, including rotating black holes. However, a significant simplification arises when the perturbed configuration is static and spherically symmetric.
For spherically symmetric configurations, the angular dependence can be separated using spherical harmonics. In this case, the angular part of Eq.~(\ref{angular}) reduces to the associated Legendre functions,
\begin{equation}
S_{\ell m}(\theta)=e^{-i m\phi}Y_{\ell m}(\theta,\phi)=P_{\ell m}(\cos\theta).
\end{equation}
The corresponding separation constants are independent of $m$, reflecting their degeneracy:
\begin{equation}\label{eq:ssseparation}
\Lambda_{\ell m}=\ell(\ell+1).
\end{equation}

It is convenient to rewrite Eq.~(\ref{schroedinger_GBF}) as
\begin{equation}\label{eq:schroedinger_simple}
\frac{d^2\Psi}{dr_*^2}+(\omega^2-V_{\ell}(r_*))\Psi(r_*)=0,
\end{equation}
where $V_{\ell}$ does not depend on $m$ and, when there is no Coulomb interaction, $V_{\ell}$ vanishes at the horizons. In the asymptotic limit,
\begin{equation}
\lim_{r\to\infty}V_{\ell}(r)=\mu^2,
\end{equation}
where $\mu$ is the field's mass.

The same construction applies to higher-dimensional spherically symmetric configurations. The generalization of spherical harmonics to $D$ dimensions leads to
\begin{equation}\label{eq:ssseparationD}
\Lambda_{\ell}=\ell(\ell+D-3),
\end{equation}
and the equation for the radial function is again reduced to the form of Eq.~(\ref{eq:schroedinger_simple}).

\section{Physical applications of grey-body factors}\label{sec:applications}

The grey-body factors defined in the scattering problem are directly connected to several physical observables. In this section we collect three representative applications: absorption cross sections, Hawking radiation, and the frequency-domain description of black-hole ringdown.

\subsection{Absorption cross section}\label{sec:absoption}
The absorption cross section represents the effective area of a black hole for capturing incoming radiation of a given frequency. Unlike grey-body factors, which describe transmission for individual partial waves, it provides a cumulative measure obtained by summing over all angular modes and therefore has a direct interpretation as an observable quantity. In this sense, it characterizes how efficiently the black hole absorbs incident flux and connects scattering properties with classical notions of capture. Because the scattering problem is naturally formulated in terms of plane waves, we first relate them to spherical waves in a $D$-dimensional spacetime and then express the absorption cross section in terms of grey-body factors.

Assuming that spherical symmetry is restored in the asymptotic region, we introduce an orthonormal basis of hyperspherical harmonics $Y_{\ell a}(\Omega)$ on the $(D-2)$-sphere, parametrized by angular coordinates $\Omega$, where $a=1,\ldots,d_\ell$ labels the degenerate basis elements at fixed $\ell$. These harmonics satisfy
\begin{equation}\label{eq:hyper_ortho}
\int_{S^{D-2}} d\Omega\, Y_{\ell a}(\Omega)\,
\overline{Y_{\ell' a'}(\Omega)}
= \delta_{\ell\ell'}\delta_{aa'},
\end{equation}
where $\ell$ denotes the multipole number, which labels the eigenvalues of the Laplacian on the $(D-2)$-dimensional sphere. The hyperspherical harmonics $Y_{\ell a}(\Omega)$ satisfy
\begin{equation}
\nabla^2_S Y_{\ell a}(\Omega) = -\ell(\ell + D - 3)\, Y_{\ell a}(\Omega),
\end{equation}
The index $a$ distinguishes the different basis elements belonging to the same eigenvalue. The total number of independent modes (hyperspherical harmonics) associated with a given value of $\ell$ is given by \cite{Bander:1965rz}
\begin{equation}\label{eq:degeneracy}
d_\ell = \frac{(2\ell + D - 3)(\ell + D - 4)!}{\ell!(D - 3)!}.
\end{equation}
In particular, for $D=4$, $d_\ell = 2\ell + 1$.

A plane wave admits an expansion in terms of Gegenbauer polynomials \cite{olfver2010nist}:
\begin{equation}
e^{\imo kz} = e^{\imo kr \cos\theta}
= \frac{\Gamma(\nu)}{(kr/2)^\nu}
\sum_{\ell=0}^{\infty} (\ell + \nu)\,\imo^\ell\,
J_{\ell+\nu}(kr)\, C_\ell^{(\nu)}(\cos\theta),
\qquad \nu \equiv \frac{D-3}{2}.
\end{equation}
Here $J_\alpha(z)$ are Bessel functions of the first kind. The Gegenbauer polynomials $C_\ell^{(\nu)}(\cos\theta)$ can be expressed in terms of hyperspherical harmonics as
\begin{equation}\label{eq:Gegenbauer_exp}
\sum_{a=1}^{d_\ell} Y_{\ell a}(\Omega)\,
\overline{Y_{\ell a}(\Omega_k)}
= \frac{d_\ell}{\Omega_{D-2}} \frac{C_\ell^{(\nu)}(\cos\theta)}{C_\ell^{(\nu)}(1)}
= \frac{\ell + \nu}{\nu\,\Omega_{D-2}}\, C_\ell^{(\nu)}(\cos\theta),
\end{equation}
where
\begin{equation}
\Omega_{D-2} = \frac{2\pi^{(D-1)/2}}{\Gamma\!\left(\frac{D-1}{2}\right)}
\end{equation}
is the area of the unit $(D-2)$-sphere, and
\begin{equation}
C_\ell^{(\nu)}(1) = \frac{(\ell + 2\nu - 1)!}{(2\nu - 1)!\,\ell!}.
\end{equation}

The angles $\Omega_k$ specify a point on the $(D-2)$-sphere corresponding to the direction of the plane wave. We choose coordinates such that $\theta$ is the angle between $\Omega$ and $\Omega_k$. In particular, when $\Omega = \Omega_k$, we have $\theta = 0$, and Eq.~(\ref{eq:Gegenbauer_exp}) reduces to
\begin{equation}
\frac{d_\ell}{\Omega_{D-2}} = \sum_{a=1}^{d_\ell} |Y_{\ell a}(\Omega_k)|^2.
\end{equation}
This relation fixes the sum over the degenerate basis, whereas the individual values of $|Y_{\ell a}(\Omega_k)|^2$ depend on the choice of basis and on the incident direction $\Omega_k$. When $\Omega_k$ is chosen to be the axial direction,
\begin{equation}
|Y_{\ell a}(\Omega_k)|^2 = \frac{1}{\Omega_{D-2}}.
\end{equation}

Substituting Eq.~(\ref{eq:Gegenbauer_exp}) into the plane-wave expansion yields
\begin{equation}\label{eq:planewave_exp}
e^{\imo kz}
=
\frac{(2\pi)^{\nu+1}}{(kr)^\nu}
\sum_{\ell=0}^{\infty} \sum_{a=1}^{d_\ell}
\imo^\ell\, J_{\ell+\nu}(kr)\,
Y_{\ell a}(\Omega)\, \overline{Y_{\ell a}(\Omega_k)}.
\end{equation}

To separate ingoing and outgoing components, we use the large-argument asymptotics of the Bessel function:
\begin{equation}\label{eq:Bessel_asymptotic}
J_{\ell+\nu}(kr)
=
\frac{1}{\sqrt{2\pi kr}}
\left(
e^{\imo\left(kr - \frac{\pi}{2}(\ell+\nu) - \frac{\pi}{4}\right)}
+
e^{-\imo\left(kr - \frac{\pi}{2}(\ell+\nu) - \frac{\pi}{4}\right)}
\right).
\end{equation}
Substituting Eq.~(\ref{eq:Bessel_asymptotic}) into Eq.~(\ref{eq:planewave_exp}), we obtain
\begin{equation}\label{eq:planewave_final}
e^{\imo kz}
=
\sum_{\ell=0}^{\infty} \sum_{a=1}^{d_\ell}
\left(
K^{\mathrm{(out)}}_{\ell a} e^{\imo kr}
+
K^{\mathrm{(in)}}_{\ell a} e^{-\imo kr}
\right)
\frac{Y_{\ell a}(\Omega)}{r^{(D-2)/2}},
\end{equation}
with coefficients
\begin{align}
K^{\mathrm{(out)}}_{\ell a}
&=
\frac{(2\pi)^{(D-2)/2}}{k^{(D-2)/2}}
\overline{Y_{\ell a}(\Omega_k)}\,
e^{-\imo\pi (D-2)/4}, \\
K^{\mathrm{(in)}}_{\ell a}
&=
\frac{(2\pi)^{(D-2)/2}}{k^{(D-2)/2}}
\overline{Y_{\ell a}(\Omega_k)}\,
(-1)^\ell\, e^{\imo\pi (D-2)/4}.
\end{align}

The ingoing component in the channel $(\ell,a)$ in the partial-wave decomposition~(\ref{eq:planewave_final}) has amplitude $K^{\mathrm{(in)}}_{\ell a}$. By the orthonormality relation~(\ref{eq:hyper_ortho}), this component carries an integrated inward radial flux $k\left|K^{\mathrm{(in)}}_{\ell a}\right|^2$, whereas the plane wave $e^{\imo kz}$ has incident flux density $k$. The flux absorbed from the channel $(\ell,a)$ is obtained by multiplying its incoming flux by $\Gamma_{\ell a}(\omega)$. Hence, the absorption cross section of the plane wave is
\begin{equation}\label{eq:absorption}
\sigma_{\mathrm{abs}}(\omega)
=
\sum_{\ell=0}^{\infty} \sum_{a=1}^{d_\ell}
\left|K^{\mathrm{(in)}}_{\ell a}\right|^2 \Gamma_{\ell a}(\omega)
=
\frac{(2\pi/k)^{D-2}}{\Omega_{D-2}}
\sum_{\ell=0}^{\infty} \sum_{a=1}^{d_\ell}
\Gamma_{\ell a}(\omega).
\end{equation}
Here we used
\begin{equation}
\left|K^{\mathrm{(in)}}_{\ell a}\right|^2
=
\left(\frac{2\pi}{k}\right)^{D-2}
\left|Y_{\ell a}(\Omega_k)\right|^2.
\end{equation}

For spherically symmetric black holes, the grey-body factors are independent of the degeneracy index $a$. Eq.~(\ref{eq:absorption}) then simplifies to
\begin{equation}\label{eq:absorption_spherical}
\sigma_{\mathrm{abs}}(\omega)
=
\frac{(2\pi/k)^{D-2}}{\Omega_{D-2}}
\sum_{\ell=0}^{\infty} d_\ell\, \Gamma_\ell(\omega)
=
\frac{(4\pi)^{(D-3)/2}\Gamma\!\left(\frac{D-1}{2}\right)}{k^{D-2}}
\sum_{\ell=0}^{\infty}
\frac{(2\ell + D - 3)(\ell + D - 4)!}{\ell!(D - 3)!}
\,\Gamma_\ell(\omega).
\end{equation}

It is important to note that, for fields with nonzero spin, the total number of independent modes associated with a given value of the separation constant does not coincide with the number of hyperspherical harmonics. Instead, one must count the physical degrees of freedom (excluding gauge redundancies), which depend on the spacetime dimension.

In a $D$-dimensional spacetime (i.e., on a $(D-2)$-sphere), the numbers of independent modes corresponding to vector and symmetric tensor degrees of freedom are, respectively, \cite{10.1063/1.526034,10.1063/1.526749}
\begin{eqnarray}\label{eq:degeneracy_vector}
d_\ell &=& \frac{\ell(\ell + D - 3)(2\ell + D - 3)(\ell + D - 5)!}{(\ell+1)!(D - 4)!}, \qquad\ell=1,2,3\ldots.
\\
\label{eq:degeneracy_tensor}
d_\ell &=& \frac{(D-4)(D-1)(\ell + D - 2)(\ell - 1)(2\ell + D - 3)(\ell + D - 5)!}{2(\ell+1)!(D - 3)!}, \qquad\ell=2,3,4\ldots.
\end{eqnarray}
These multiplicities account only for modes transforming as vectors or tensors under rotations on the $(D-2)$-sphere. Such modes are typically governed by the same master equation (share the same effective potential) and can therefore be treated collectively. For bosonic fields of spin $1$, such as the Maxwell field, the degrees of freedom decompose into components that transform as scalars and as vectors on the $(D-2)$-sphere. Similarly, gravitational perturbations split into scalar-, vector-, and tensor-type components \cite{Ishibashi:2011ws}. From Eq.~(\ref{eq:degeneracy_tensor}), one finds that in $D=4$ the number of tensor-type degrees of freedom vanishes, indicating that these modes are pure gauge. In this case, the gravitational perturbations are classified into vector-type (odd-parity) and scalar-type (even-parity) sectors \cite{Regge:1957td}, each with degeneracy $d_\ell = 2\ell + 1$.

\subsection{Hawking radiation}\label{sec:Hawking}

Grey-body factors play a central role in determining the observable spectrum of Hawking radiation. Although the Hawking effect is often characterized by a thermal spectrum at the black-hole temperature, the radiation detected at infinity is not a perfect blackbody. Particles created near the horizon must propagate through the curvature-induced effective potential surrounding the black hole, and only a fraction of each mode is transmitted to the asymptotic region. The corresponding transmission probabilities, or grey-body factors, therefore modify both the spectral distribution and the total luminosity of the emitted radiation, making them indispensable in quantitative studies of black-hole evaporation.

Black-hole evaporation is a gradual process whose character changes significantly as the mass decreases and the Hawking temperature increases. For sufficiently large black holes, the radiation is extremely weak, and the evolution is well described within the semiclassical approximation. As the black hole loses mass, however, additional particle species become available for emission, the evaporation rate accelerates, and eventually the process enters a regime where quantum-gravity effects are expected to become important. Schematically, one may distinguish the following stages.

\begin{enumerate}
\item \emph{Semiclassical Hawking-radiation regime.}
The evaporation begins with Hawking radiation whose characteristic temperature is determined by the surface gravity of the event horizon \cite{Hawking:1975vcx},
\begin{equation}
T_H=\frac{\kappa}{2\pi} = \frac{F'(r_H)}{4\pi}.
\end{equation}
For the four-dimensional Schwarzschild metric, $F(r)=1-2M/r$, where $M$ is the black-hole mass, and the temperature reduces to
\begin{equation}
T_H=\frac{1}{8\pi M}.
\end{equation}
Massive black holes evaporate negligibly slowly on astrophysical timescales. For example, a black hole of one solar mass has a temperature of only
\begin{equation}
T_H \approx 6\times10^{-8}\ {\rm K}.
\end{equation}
The spectrum measured at infinity is filtered by the effective potential barrier and has the form
\begin{equation}
\frac{\partial^2E}{\partial t\partial\omega}
=
\sum_{\ell,m}
\frac{\Gamma_{\ell m}(\omega)}{2\pi}
\frac{\omega}
{\exp\left[(\omega-\Phi_H)/T_H\right]\pm1},
\end{equation}
where plus sign in the denominator corresponds to fermions and the minus sign to bosons. In practical evaporation calculations, this expression is only the contribution of a given particle species and set of internal degrees of freedom. The total primary spectrum must be obtained by summing over all particles that can be emitted, with their spin, mass threshold, statistics, degeneracy, and species-dependent grey-body factors taken into account. Massive fields require grey-body factors that incorporate both the propagation threshold and the polarization structure. For Kerr black holes, recent calculations have determined the Hawking emission of massive vector fields, including superradiant modes and the corresponding rates of mass and angular-momentum loss \cite{Calza:2026dpr}. During this stage, the black-hole mass decreases slowly, while the Hawking temperature gradually rises.

\item \emph{Emission of light particles.}
As the temperature increases, progressively heavier particle species become kinematically accessible. At low temperatures the emission is dominated by massless or nearly massless particles, such as photons, gravitons, and neutrinos. Once the Hawking temperature becomes comparable to the electron mass,
\begin{equation}
T_H \gtrsim m_e \approx 0.5MeV\approx 0.58\cdot10^{10}K,
\end{equation}
the black hole begins to emit electron--positron pairs efficiently. Higher temperatures subsequently open muon, hadronic, and other Standard Model channels, increasing the number of radiating degrees of freedom and accelerating the mass loss.

\item \emph{The QCD regime.}
When the Hawking temperature approaches the characteristic scale of strong interactions,
\begin{equation}
T_H \sim \Lambda_{\rm QCD}
\approx 200~{\rm MeV}\approx 2.3\cdot10^{12}K,
\end{equation}
the black hole begins to emit quarks and gluons directly. These particles cannot propagate freely and quickly hadronize into jets containing pions, kaons, nucleons, and high-energy photons \cite{MacGibbon:1990zk,MacGibbon:1991vc}. At this stage the observed spectrum is no longer a simple thermal distribution. Secondary decays and hadronization processes produce a rich particle cascade, making the phenomenology substantially more complicated than in the low-temperature regime. This phase is particularly important in studies of primordial black holes whose masses are sufficiently small for such temperatures to be reached during the present epoch \cite{Auffinger:2022khh}.

\item \emph{The electroweak regime.}
As the evaporation proceeds further and the temperature reaches the electroweak and heavy-particle mass scales,
\begin{equation}
T_H \gtrsim 100~{\rm GeV}\approx 1.16\cdot10^{15}K,
\end{equation}
all known Standard Model degrees of freedom can contribute, and their large multiplicity rapidly increases the luminosity while shortening the remaining lifetime.

\item \emph{The quantum-gravity regime.}
Eventually the black-hole mass approaches the Planck scale,
\begin{equation}
M \sim M_{\rm Pl},
\end{equation}
or, equivalently, the horizon radius becomes comparable to the Planck length. In this regime the semiclassical description is no longer reliable, and a complete theory of quantum gravity is required. At present there is no consensus regarding the final outcome of the evaporation process. In the simplest scenario the black hole evaporates completely, leaving no remnant behind \cite{Hawking:1976ra}. Another possibility is the formation of a stable or long-lived Planck-scale remnant. Such remnants arise naturally in many regular black-hole models, noncommutative geometries, generalized uncertainty principle scenarios, and other quantum-gravity-inspired constructions. A third possibility is a final burst of energetic particles that releases the remaining mass in a short-lived explosive event.
\end{enumerate}

The evaporation of regular black holes often differs qualitatively from that of the Schwarzschild solution because the modification that removes the central singularity commonly produces an inner horizon in addition to the outer event horizon. As the mass decreases, the two horizons approach each other and merge in an extremal configuration at a minimum mass. At this point the metric function has a double zero, so the surface gravity and the Hawking temperature vanish \cite{Banks:1992ba}. The evaporation therefore proceeds through three characteristic stages:
\begin{enumerate}
\item At large masses, the black hole behaves similarly to the Schwarzschild solution and the temperature increases during evaporation.
\item As the mass approaches a critical value, the temperature attains its maximum and the evaporation rate becomes most efficient.
\item Near the extremal regime, the temperature decreases and eventually tends to zero,
\begin{equation}
T_H \to 0.
\end{equation}
The evaporation then slows down dramatically, asymptotically approaching a cold remnant rather than ending in complete disappearance.
\end{enumerate}
This behavior is encountered in a wide variety of regular and quantum-corrected black-hole models, including Bardeen, Hayward, noncommutative, and several Einstein--Yang--Mills solutions \cite{Bolokhov:2024voa,Hayward:2005gi,Nicolini:2005vd,Ansoldi:2008jw,Bonanno:2000ep,Bonanno:2006eu,Modesto:2004xx,Balart:2014cga,Flachi:2012nv}. In such cases the late stages of evaporation are particularly interesting, since both the Hawking temperature and the grey-body factors may differ substantially from their Schwarzschild counterparts, leading to distinctive emission spectra and potentially observable signatures. Nonminimal scalar-curvature coupling can also substantially enhance or suppress the Hawking spectra of several regular black-hole models \cite{Calza:2025yfm}. Recent calculations have quantified the role of grey-body factors in scalar emission from rotating black-bounce geometries \cite{Calza:2022ioe}.

\subsection{Grey-body factors and black-hole ringdown}\label{sec:ringdown_gbf}

The original idea that grey-body factors can be inferred directly from the ringdown waveform was proposed in Ref.~\cite{Oshita:2023cjz}. In this approach the post-merger signal is studied in the frequency domain, where the curvature-induced potential barrier around the remnant acts as a filter for perturbations generated during the merger. The standard quasinormal-mode description emphasizes the poles of the retarded Green function, while the grey-body-factor description emphasizes the real-frequency scattering response of the same geometry. The observed ringdown is therefore not determined solely by the quasinormal frequencies: its spectral envelope also carries information about how efficiently each frequency is reflected or transmitted by the effective potential.

For the dominant gravitational-wave multipoles, early studies found that the ringdown amplitude can be described by a simple function of the grey-body factor, or equivalently by the modulus of the reflection amplitude of the remnant. This relation was first developed for the ringdown of a Kerr black hole produced in an extreme-mass-ratio merger \cite{Oshita:2023cjz}, and was later extended to merging binary black holes \cite{Okabayashi:2024qbz}. The frequency-domain amplitude model was subsequently refined for comparable-mass mergers in Ref.~\cite{Rosato:2025ulx}. The physical interpretation is straightforward: frequencies for which the potential barrier is mostly reflective leave a strong imprint in the reflected component of the post-merger signal, while frequencies that are efficiently transmitted are comparatively suppressed in that component. Since the height, width, and location of the barrier depend on the remnant parameters, the shape of the ringdown spectrum contains information about the mass and spin of the final black hole.

This first formulation mainly used the grey-body factor as an amplitude filter. It captured the dominant spectral envelope, but it did not make full use of the complex scattering information. In particular, the reflection amplitude also has a phase, and this phase contributes to the frequency-domain phase of the waveform. Moreover, the observed signal depends not only on the scattering properties of the remnant, but also on how the remnant is excited by the nonlinear merger. These considerations motivate a more complete model in which the source-dependent excitation and the geometry-dependent scattering response are separated as much as possible.

Recent work has made this separation more quantitative. In particular, Ref.~\cite{Rosato:2026moe} showed that the reflection amplitude is directly encoded in the component of the Green function controlling the linear ringdown response in the frequency domain. Building on this observation, Ref.~\cite{Rosato:2026apq} introduced \texttt{GreyRing}, a theory-informed model in which each frequency-domain multipole of the post-merger signal is written as
\begin{equation}\label{eq:greyring_model}
\widetilde h_{\ell m}(\omega) = \frac{M A_{\ell m}}{(M\omega)^{p_{\ell m}}} R_{\ell m} \exp\left(\frac{\imo c_{\ell m}}{M\omega}\right) \exp\left[\imo(\phi_c+\omega t_c)\right].
\end{equation}
Here \(M\) is the remnant mass, and \(R_{\ell m}\) is the complex reflection amplitude of the remnant geometry discussed in Sec.~\ref{sec:scattering}. The parameters \(t_c\) and \(\phi_c\) are the time and phase of coalescence, while \(A_{\ell m}\), \(p_{\ell m}\), and \(c_{\ell m}\) are phenomenological parameters that encode the source-dependent excitation and the residual frequency dependence not contained in \(R_{\ell m}\). The important separation is that the source parameters determine the overall envelope and phase correction, while the mass and spin of the final black hole enter through the grey-body-factor scattering data. The grey-body factor controls the spectral amplitude through \(|R_{\ell m}|\), while the full complex reflection amplitude also contributes a frequency-dependent scattering phase. The modulus alone is sufficient to understand why the signal is suppressed or enhanced in different frequency ranges, but the phase of \(R_{\ell m}\) is needed for an accurate waveform model.

This point is particularly relevant for black-hole spectroscopy. Standard spectroscopy attempts to measure several quasinormal frequencies and check whether they correspond to the same remnant mass and spin. In practice, however, the result may depend on the chosen ringdown start time and on how many overtones are included. A grey-body-factor model instead uses the continuous real-frequency shape of the post-merger spectrum. In this representation, changing \(M\) rescales the argument \(M\omega\), while changing the rotation parameter changes the spin-dependent scattering response of the final Kerr black hole. Ref.~\cite{Rosato:2026apq} showed that this strategy can reproduce a large set of comparable-mass, aligned-spin numerical-relativity waveforms, with typical mismatches of order \(10^{-6}\) for the dominant \((\ell,m)=(2,2)\) mode. The same work proposed a consistency test in which the mass and angular momentum inferred from the grey-body-factor waveform are compared with the values obtained from standard black-hole spectroscopy or from the full inspiral-merger-ringdown signal.

Several limitations should be kept in mind. The grey-body-factor imprint does not determine the waveform by itself: the excitation of the remnant still depends on the nonlinear merger dynamics and is represented in Eq.~(\ref{eq:greyring_model}) by phenomenological source-dependent parameters. In data analysis one must also select a frequency interval in which the post-merger model is valid: a lower cutoff that is too low retains inspiral contamination, whereas a cutoff that is too high leaves too little signal and becomes noise dominated. Moreover, higher multipoles, mode mixing, precession, eccentricity, and possible deviations from Kerr may require extensions of the current templates. Nevertheless, the emerging picture is clear: grey-body factors provide a direct bridge between black-hole scattering theory and ringdown observations, complementing the quasinormal-mode description and opening a new way to test the strong-field geometry of the remnant.

\section{Analytic and semi-analytic methods}\label{sec:analytic}

Exact closed-form grey-body factors are known only in special limits, so analytic and semi-analytic methods play a complementary role to direct numerical integration. They identify which geometric features of the effective potential control transmission, provide useful checks on numerical calculations, and yield compact formulas in low-frequency, high-frequency, eikonal, or single-barrier regimes. The methods reviewed in this section should therefore be understood not as mutually exclusive alternatives, but as approximations with different domains of validity and different degrees of physical transparency.

\subsection{WKB approach for grey-body factors}\label{sec:WKB}

One of the most efficient and economical methods for finding grey-body factors is the Wentzel–-Kramers–-Brillouin (WKB) method \cite{Schutz:1985km,Iyer:1986nq}. The main idea of the method is that the solution of the wave equation is expanded in a Taylor series near the peak of the effective potential and then matched to the WKB expansions at the asymptotic boundaries. Consequently, for the method to be valid, the effective potential must possess a single peak and decay sufficiently slowly at the boundaries, so that the WKB expansion can be applied.

\begin{figure}[t]
\centering
\includegraphics[width=\linewidth]{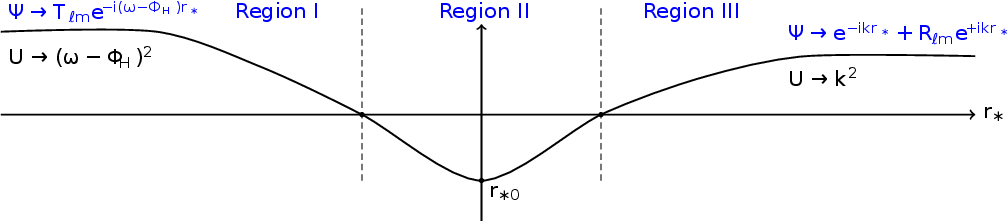}
\caption{Schematic illustration of the WKB treatment of barrier scattering in terms of the tortoise coordinate $r_*$. The function $U$ possesses a single minimum, giving rise to two turning points which separate three regions: Region I (near the horizon), Region II (near the peak), and Region III (asymptotic infinity). The boundary conditions appropriate for grey-body factors are imposed as an incident and reflected wave at spatial infinity, and a purely transmitted wave at the horizon. The WKB method matches solutions across the barrier via a uniform approximation in Region II.}
\label{fig:wkb_regions}
\end{figure}

We assume that the function $U$ in the master equation~(\ref{schroedinger_GBF}) has a single minimum, which corresponds to the maximum of the effective potential $V_{\ell}$ in Eq.~(\ref{eq:schroedinger_simple}) at $r_{*}=r_{*0}$, where
$$U'(r_{*0})=0, \qquad U''(r_{*0})>0.$$
The essential idea is to treat the region near the maximum separately from the asymptotic regions, where the usual WKB solutions are valid (see Fig.~\ref{fig:wkb_regions}).
In the vicinity of the peak, the function $U(r_{*})$ is expanded as a Taylor series,
\begin{equation}\label{eq:potexpansion}
U(r_{*})=U_0+\frac{1}{2}U_2(r_{*}-r_{*0})^2+\frac{1}{6}U_3(r_{*}-r_{*0})^3+\cdots,
\end{equation}
where all derivatives are evaluated at $r_{*0}$. The crucial step is then to introduce a rescaled coordinate,
\begin{equation}
z=(2U_2)^{1/4}e^{\imo\pi/4}(r_{*}-r_{*0}),
\end{equation}
such that the leading part of the equation reduces to a \emph{parabolic cylinder equation}. To lowest nontrivial order, one obtains
\begin{equation}
\frac{d^2\Psi}{dz^2}+\left(\K-\frac{z^2}{4}\right)\Psi=0,\qquad \K\equiv\imo\frac{U_0}{\sqrt{2U_2}},
\end{equation}
whose general solution can be expressed in terms of parabolic cylinder functions,
\begin{equation}\label{eq:uniform}
\Psi(z)=A D_{\K-1/2}(z)+B D_{-\K-1/2}(\imo z).
\end{equation}
Here $D_\nu(z)$ is formally defined as the unique solution of Weber's equation
$$
\frac{d^2D_\nu(z)}{dz^2}+\left(\nu+\frac{1}{2}-\frac{z^2}{4}\right)D_\nu(z)=0
$$
with the standard normalization
$$
D_\nu(0)=\frac{2^{\nu/2}\sqrt{\pi}}{\Gamma\!\left(1/2-\nu/2\right)},
\qquad
D_\nu'(0)=-\frac{2^{(\nu+1)/2}\sqrt{\pi}}{\Gamma\!\left(-\nu/2\right)}.
$$

Thus, instead of matching WKB solutions across two separate turning points, one constructs a \emph{uniform approximation} in the barrier region (\ref{eq:uniform}), and matches this solution to the WKB solutions on both sides simultaneously.

In this way, by comparing the asymptotic expansions for (\ref{eq:uniform}) with (\ref{eq:scattering}), one can express the coefficients $R_{\ell m}$ and $T_{\ell m}$, as well as $A$ and $B$, in terms of $\K$.

At leading order one finds that the transmission and reflection coefficients are given by
\begin{eqnarray}\label{eq:reflection}
|R|^2 &=& \frac{1}{1+e^{-2\pi\imo \K}}, \qquad 0<|R|^2<1,\\
\label{eq:transmission}
|T|^2 &=& \frac{1}{1+e^{2\pi\imo \K}} = 1-|R|^2.
\end{eqnarray}

Importantly, $\K$ is not, in general, restricted to its lowest-order value but is instead obtained by solving an algebraic relation involving $U_0$ and its derivatives at higher orders. Finally, one obtains \cite{Iyer:1986np}
\begin{equation}\label{eq:WKBexpansion}
\imo\frac{U_0}{\sqrt{2U_2}} - \Lambda_{2}(\K) - \Lambda_{3}(\K) \ldots = \K,
\end{equation}
where $\Lambda_i(\K)$ can be expressed in terms of $U_2,U_3,\ldots,U_{2i}$. The WKB calculation can be extended to higher orders, as shown in \cite{Konoplya:2003ii,Matyjasek:2017psv,Matyjasek:2019eeu,Matyjasek:2026yiu,Konoplya:2026rjh}. WKB orders up to the sixth usually provide the best accuracy for grey-body factors.

Thus, the practical procedure for computing grey-body factors is as follows:
\begin{enumerate}
\item For a given real frequency $\omega$, determine the maximum of the effective potential and compute $U_0$, $U_2$, and higher derivatives.
\item
Solve Eq.~(\ref{eq:WKBexpansion}) with respect to $\K$.
\item Substitute $\K$ into (\ref{eq:transmission}) to calculate the grey-body factor
\begin{equation}\label{eq:gb_wkb}
\Gamma_{\ell m}(\omega)=|T|^2=\frac{1}{1+e^{2\pi\imo \K}}.
\end{equation}
\end{enumerate}

The use of parabolic cylinder functions ensures that the approximation remains valid even when the turning points coalesce near the peak, which is precisely the regime relevant for grey-body factors at moderate frequencies. Higher-order corrections systematically improve the accuracy by incorporating higher derivatives of the potential, allowing the method to extend well beyond the eikonal limit and to remain reliable even for relatively small multipole numbers. The above WKB method at various orders has been applied to compute grey-body factors of black holes, wormholes, and black branes in a number of publications \cite{Konoplya:2019ppy,Konoplya:2021ube,Konoplya:2023ppx,Malik:2025czt,Liang:2026eic,Lutfuoglu:2026uzy}, and has usually shown reasonable accuracy for $\ell > s$, where $s$ is the spin of the field.

\subsection{Eikonal approximation and its relation to the WKB method}\label{sec:EikonalWKB}

It is important to distinguish the WKB approximation from the eikonal approximation, although the two are closely related. The eikonal approximation corresponds to the short-wavelength or geometric-optics regime, in which the multipole number is large,
\begin{equation}
\ell \gg 1, \qquad \Lambda_{\ell}\gg1.
\end{equation}
In this limit the angular momentum term dominates the effective potential, and the wave propagation becomes closely related to the motion of classical null particles in the corresponding spacetime geometry.

For asymptotically flat spherically symmetric black holes, the effective potential in (\ref{eq:schroedinger_simple}) typically takes the form
\begin{equation}\label{eq:effpot_eikonal}
V_{\ell}(r)=f(r)\frac{\Lambda_{\ell}}{r^2}+\Order{1},
\end{equation}
where $f(r)$ is the lapse function. The maximum of the potential is therefore determined by the condition
\begin{equation}\label{eq:photon_sphere_condition}
\frac{d}{dr}\frac{f(r)}{r^2}=0,
\end{equation}
which coincides precisely with the condition for the circular null orbit (\ref{eq:photon-sphere-condition}), i.e.~the photon sphere radius $r=r_c$.

Thus, in the eikonal limit the scattering problem is governed by the properties of null geodesics near the photon sphere. Waves with frequencies close to the height of the potential barrier spend a long time orbiting near the unstable circular trajectory before either escaping to infinity or falling into the black hole. This geometric interpretation explains the close connection between grey-body factors, absorption cross sections, and quasinormal modes in the high-frequency regime (see Sec.~\ref{sec:correspondence}).

The WKB approximation becomes asymptotically exact in the eikonal limit. Indeed, when $\Lambda_{\ell}$ is large, the effective potential varies slowly compared to the local wavelength, and the dominant contribution to the scattering coefficients is determined only by the behavior of the potential near its maximum. Retaining only the leading WKB order in Eq.~(\ref{eq:transmission}), one obtains
\begin{equation}\label{eq:gb_eikobal}
\Gamma_{\ell}(\omega)\approx
\frac{1}{1+\exp\!\left(
-2\pi \frac{\omega^2-V_{\ell}(r_{*0})}{\sqrt{-2V_{\ell}''(r_{*0})}}
\right)},
\end{equation}
where $V_{\ell}(r_{*0})$ and $V_{\ell}''(r_{*0})$ denote the value of the effective potential and its second derivative with respect to the tortoise coordinate at the maximum ($r=r_c$). This expression shows that the transmission probability rapidly changes from nearly zero to nearly unity as the frequency crosses the barrier height.

In the high-frequency regime the grey-body factors approach a step-like profile,
\begin{equation}
\Gamma_{\ell}(\omega)\to
\Theta(\omega^2-V_{\ell}(r_c)),
\end{equation}
where $\Theta$ is the Heaviside function.

Although the eikonal approximation provides an important analytical description of black-hole scattering, it should not be identified with the WKB method itself. Higher-order WKB expansions remain applicable beyond the strict eikonal regime and often provide accurate grey-body factors even for moderate multipole numbers \cite{Konoplya:2023moy,Dubinsky:2024nzo}. Nevertheless, the accuracy deteriorates for low multipoles, especially for $\ell=0$, or for frequencies far below the peak of the potential barrier. The standard WKB construction also becomes unreliable when the effective potential possesses multiple extrema, discontinuities, or rapidly varying structure, because the matching procedure between turning points is then no longer uniformly valid.

Intuitively, one might expect that the eikonal regime $\ell\to\infty$ always allows one to apply the WKB formula (\ref{eq:gb_eikobal}), since the centrifugal term (\ref{eq:effpot_eikonal}) dominates the effective potential. However, the situation is more subtle. In a number of theories, particularly those involving higher-curvature corrections, the centrifugal term for gravitational perturbations may deviate from the standard form (\ref{eq:effpot_eikonal}), leading even to a catastrophic instability of the background metric \cite{Dotti:2004sh,Dotti:2005sq,Gleiser:2005ra,Konoplya:2008ix,Takahashi:2010gz,Takahashi:2011qda,Cuyubamba:2016cug,Konoplya:2017ymp,Konoplya:2019hml,Konoplya:2020bxa}, or the metric function $f(r)$ may be such that the effective potential develops a double-well structure \cite{Guo:2022ghl,Konoplya:2025hgp}. As a result, the WKB method may fail to provide reliable results even for very large multipole numbers \cite{Konoplya:2017wot,Konoplya:2019hml}. Additional examples of the breakdown of the WKB approximation can be found in \cite{Bolokhov:2023dxq,Konoplya:2025afm}.

\subsection{Low-energy absorption cross section and its universality}\label{sec:lowfrequencymatching}

In the seminal work~\cite{Das:1996we}, the low-energy absorption cross section for massless fields propagating in a general spherically symmetric black hole background was investigated. The main result of that paper was the demonstration of a remarkable universality: for a minimally coupled scalar field, the absorption cross section in the limit $\omega \to 0$ coincides exactly with the area of the event horizon.

We consider a $D$-dimensional spherically symmetric black hole with metric
\begin{equation}\label{eq:higherdimensionalmetric}
ds^2 = -f(r)\,dt^2 + \frac{f(r)}{F^2(r)}\,dr^2 + r^2 d\Omega_{D-2}^2,
\end{equation}
where $d\Omega_{D-2}^2$ is the line element on the unit $(D-2)$-sphere.
In an asymptotically flat spacetime, the function $F(r)$ behaves as
\begin{equation}
F(r)=1-\frac{r_H^{D-3}}{r^{D-3}}+\order{\frac{1}{r^{D-2}}},
\qquad
\lim_{r\to\infty}\frac{r_*}{r}=1.
\end{equation}

Therefore, for a massless scalar field, the effective potential takes the form
\begin{equation}
V_{\ell}(r)=\frac{4\Lambda_{\ell}+(D-2)(D-4)}{4r^2}+\order{\frac{1}{r^2}}=\frac{4\Lambda_{\ell}+(D-2)(D-4)}{4r_*^2}+\order{\frac{1}{r_*^2}},
\end{equation}
where $\Lambda_{\ell}$ is the separation constant defined in Eq.~(\ref{eq:ssseparationD}). For $\ell=0$, one has $\Lambda_0=0$, and Eq.~(\ref{eq:schroedinger_simple}) reduces asymptotically to
\begin{equation}\label{eq:schroedinger_simple_asymptotic}
\frac{d^2\Psi}{dr^2}+\left(\omega^2-\frac{(D-2)(D-4)}{4r^2}\right)\Psi=0.
\end{equation}

The general solution of Eq.~(\ref{eq:schroedinger_simple_asymptotic}) is proportional to a linear combination of Bessel functions,
\begin{equation}\label{eq:schroedinger_asymptotic_solution}
\Psi(r)=C_+\sqrt{r}J_{\frac{D-3}{2}}(\omega r)+C_-\sqrt{r}J_{-\frac{D-3}{2}}(\omega r).
\end{equation}

In the low-frequency limit $\omega \to 0$, and in the intermediate region $r_H \ll r \ll \omega^{-1}$, this solution behaves as
\begin{equation}\label{eq:solsmallr}
\Psi(r)=2^{-(D-3)/2}(\omega r)^{(D-2)/2}
\left[
\frac{C_+}{\Gamma\left(\frac{D-1}{2}\right)}
+\frac{2^{D-3}C_-}{(\omega r)^{D-3}\Gamma\left(\frac{5-D}{2}\right)}
\right].
\end{equation}

This expression can be matched to the near-horizon purely ingoing solution. To facilitate the matching procedure, following~\cite{Das:1996we}, we introduce the dimensionless variable
\begin{equation}
\tau=\int\frac{dr}{rF(r)}.
\end{equation}
Near the horizon, this behaves as
\begin{equation}
\tau=\frac{r_*}{r_H}+\order{r-r_H}.
\end{equation}

In terms of $\tau$, Eq.~(\ref{eq:schroedinger_simple}) reduces near the horizon to
\begin{equation}
\frac{d^2\Psi}{d\tau^2}+\omega^2 r_H^2 \Psi=0,
\end{equation}
whose purely ingoing solution is
\begin{equation}\label{eq:ingoing_approximation}
\Psi=A_{\text{tran}} e^{-i\omega r_H \tau}
= A_{\text{tran}} e^{-i\omega r_*}\left(1+\Order{r-r_H}\right).
\end{equation}

For $r \gg r_H$, the variable $\tau$ admits the expansion
\begin{equation}\label{eq:tauexpansion}
\tau=\int\frac{dr}{rF(r)}
=\ln\frac{r}{r_H}
-\frac{r_H^{D-3}}{(D-3)r^{D-3}}
+\order{\frac{1}{r^{D-3}}}.
\end{equation}
Substituting Eq.~(\ref{eq:tauexpansion}) into Eq.~(\ref{eq:ingoing_approximation}), we obtain
\begin{equation}\label{eq:ingoing_expand}
\Psi(r)=A_{\text{tran}}
\left(\frac{r}{r_H}\right)^{-i\omega r_H}
\left[
1+\frac{i\omega r_H^{D-2}}{(D-3)r^{D-3}}
+\order{\frac{1}{r^{D-3}}}
\right].
\end{equation}

Matching Eqs.~(\ref{eq:schroedinger_asymptotic_solution}) and~(\ref{eq:ingoing_expand}) in the overlap region yields~\cite{Das:1996we}
\begin{equation}
\frac{C_-}{C_+}
=i\frac{(\omega r_H)^{D-2}\Gamma\left(\frac{5-D}{2}\right)}
{2^{D-3}(D-3)\Gamma\left(\frac{D-1}{2}\right)}.
\end{equation}

As $r \to \infty$, the solution~(\ref{eq:schroedinger_asymptotic_solution}) behaves as
\begin{equation}
\Psi(r)=\sqrt{\frac{2}{\pi\omega}}
\left(
C_+\cos\left(\omega r-\frac{\pi(D-2)}{2}\right)+C_-\cos\left(\omega r+\frac{\pi(D-4)}{2}\right)
\right)
= A_{\text{in}} e^{-i\omega r}+A_{\text{out}} e^{i\omega r}.
\end{equation}

The grey-body factor for the s-wave ($\ell=0$) then reads
\begin{equation}\label{eq:swave_greybody}
\Gamma_{00}(\omega)=1-\left|\frac{A_{out}}{A_{in}}\right|^2=1-\left|\frac{1+\dfrac{C_-}{C_+}e^{\imo\pi(D-3)/2}}{1+\dfrac{C_-}{C_+}e^{-\imo\pi(D-3)/2}}\right|^2=1-\left|\frac{1+\imo\frac{(\omega r_H)^{D-2}\Gamma\left(\frac{5-D}{2}\right)}{2^{D-3}(D-3)\Gamma\left(\frac{D-1}{2}\right)}e^{\imo\pi(D-3)/2}}{1+\imo\frac{(\omega r_H)^{D-2}\Gamma\left(\frac{5-D}{2}\right)}{2^{D-3}(D-3)\Gamma\left(\frac{D-1}{2}\right)}e^{-\imo\pi(D-3)/2}}\right|^2.
\end{equation}

In the low-frequency limit $\omega r_H \ll 1$, Eq.~(\ref{eq:swave_greybody}) simplifies to
\begin{equation}\label{eq:lowfrequencygamma}
\Gamma_{00}(\omega)=
\frac{\pi(\omega r_H)^{D-2}}
{2^{D-4}\Gamma\left(\frac{D-1}{2}\right)^2}.
\end{equation}

Remarkably, Eq.~(\ref{eq:lowfrequencygamma}) is independent of the near-horizon geometry. Because the $\ell=0$ mode dominates absorption in the low-frequency regime $\omega r_H \ll 1$, Eq.~(\ref{eq:absorption_spherical}) gives
\begin{equation}\label{eq:dominant_absoroption}
\lim_{\omega\to0}\sigma_{\text{abs}}(\omega)=\lim_{\omega\to0}\frac{(4\pi)^{(D-3)/2}\Gamma\left(\frac{D-1}{2}\right)}{\omega^{D-2}}\Gamma_{00}(\omega)=A_H,
\end{equation}
where
\begin{equation}
A_H=\Omega_{D-2}r_H^{D-2}
=\frac{2\pi^{(D-1)/2}}{\Gamma\left(\frac{D-1}{2}\right)}r_H^{D-2}
\end{equation}
is the area of the event horizon. The universal low-frequency limit was verified numerically in \cite{Crispino:2009ki} for the Reissner-Nordström black hole.

An analogous analysis for spin-$1/2$ fields shows that the absorption cross section is again universal, though expressed in terms of a conformally rescaled horizon area, reflecting the different coupling of fermions to the geometry.

For vector and tensor fields, the absorption cross section vanishes in the low-frequency limit \cite{Crispino:2010fd}.

The essential physical reason behind this universality is that, in the low-energy limit, the wavelength of the incoming radiation is much larger than the black hole size, $\omega^{-1} \gg r_H$. As a result, the wave does not resolve the structure of the effective potential barrier and interacts only with the near-horizon region. Consequently, the absorption cross section becomes a purely geometric quantity determined entirely by the horizon area, independently of the detailed form of the metric functions. This establishes a clear dichotomy with the high-frequency regime, where absorption is governed instead by the photon sphere and the associated unstable null geodesics.

A similar approach was used in \cite{Kanti:2002ge} to calculate the grey-body factors and absorption cross sections for fields of nonzero spin in the background of a higher-dimensional Schwarzschild-like black hole projected onto the brane. More generally, the effective potential contains an additional term proportional to $r^{-1}$, and Eq.~(\ref{eq:ssseparationD}) can be written asymptotically as
\begin{equation}\label{eq:schroedinger_general_asymptotic}
\frac{d^2\Psi}{dr^2}+\left(\omega^2+\frac{\imo\omega\alpha}{r}+\frac{\beta}{r^2}\right)\Psi=0,
\end{equation}
where $\alpha$ and $\beta$ are dimensionless constants. Equation~(\ref{eq:schroedinger_general_asymptotic}) can then be reduced to a confluent hypergeometric equation by introducing the new variable $z=2\imo\omega r$. Its general solution is proportional to a linear combination of Kummer functions. The relation between the coefficients is determined by matching the solution in the intermediate region $r_H\ll r\ll\omega^{-1}$ (where both the near-horizon solution and the far-field solution are simultaneously valid) to the purely ingoing wave at the horizon,
\begin{equation}
\Psi(r\to r_H)=A_{\text{tran}} e^{-i\omega r_H y},
\end{equation}
expressed in terms of the appropriate variable
\begin{equation}
y=\frac{r_*}{r_H}+\Order{r-r_H},
\end{equation}
which leads to the corresponding asymptotic form for $r\gg r_H$.
As a result, the coefficients of the asymptotic solution at infinity take the form
\begin{equation}
\Psi(r\to\infty) = A_{\text{in}} e^{-i\omega r}+A_{\text{out}} e^{i\omega r}.
\end{equation}
The coefficients are expressed in terms of $A_{\text{tran}}$, which determines the reflection and transmission amplitudes and therefore the grey-body factor.

The same method has been applied to bulk emission. In the scalar-field case, the far-field equation again reduces to a Bessel equation, whereas in the general case it takes the form of a confluent hypergeometric equation. The main advantage of this approach is that it provides fully analytic expressions for the absorption probability in the regime $\omega\ll r_H^{-1}$, making explicit its dependence on the number of extra dimensions, the spin of the field, and the partial-wave number (see \cite{Kanti:2004nr} for a review). This method has therefore been widely used in studies of Hawking radiation and evaporation of higher-dimensional black holes \cite{Grain:2005my,Creek:2007sy,Creek:2007tw,Kanti:2014dxa,Antoniou:2025bvg}.

The situation is qualitatively different in asymptotically de Sitter spacetimes. The scattering problem is then posed in the finite static patch between the event horizon \(r_H\) and the cosmological horizon \(r_C\), rather than between the event horizon and spatial infinity \cite{Kanti:2005ja}. For a minimally coupled scalar field, the low-frequency \(\ell=0\) radial equation can be solved in the intermediate region between the two horizons and matched to plane waves at \(r_H\) and \(r_C\) \cite{deCesare:2025ccs}. This gives
\begin{equation}
\left|R_{00}\right|^2
=\left(\frac{r_C^2-r_H^2}{r_C^2+r_H^2}\right)^2,
\qquad
\Gamma_{00}(0)
=1-\left|R_{00}\right|^2
=\frac{4r_C^2r_H^2}{\left(r_C^2+r_H^2\right)^2}.
\end{equation}
Thus the zero-frequency transmission probability is finite whenever the cosmological horizon is located at finite radius. The dependence on the number of extra dimensions and on the cosmological constant enters through the relation between \(r_H\) and \(r_C\); increasing the cosmological constant decreases \(r_C\) and enhances the low-frequency transmission, whereas the asymptotically flat limit \(r_C\to\infty\) sends \(\Gamma_{00}(0)\) to zero.

This result changes the interpretation of the low-energy absorption cross section. If one formally uses the partial-wave normalization of Eq.~(\ref{eq:dominant_absoroption}), then the \(s\)-wave contribution behaves as
\begin{equation}
\sigma_{\rm abs}^{(\ell=0)}(\omega)
\simeq
\frac{(4\pi)^{(D-3)/2}\Gamma\left(\frac{D-1}{2}\right)}{\omega^{D-2}}
\frac{4r_C^2r_H^2}{\left(r_C^2+r_H^2\right)^2},
\qquad \omega\to0.
\end{equation}
The cross section therefore diverges instead of approaching the finite constant \(A_H\), equal to the area of the event horizon. This does not contradict the asymptotically flat universality theorem, because the assumptions behind that theorem include scattering from spatial infinity and an incident plane-wave flux normalized there. In de Sitter space the relevant quantity is more naturally the finite transmission probability between the two horizons, while the formal cross section inherits a low-frequency divergence from the \(\omega^{-(D-2)}\) prefactor because \(\Gamma_{00}(0)\) is nonzero.

Nonminimal coupling changes this low-frequency behavior. For any fixed nonzero coupling, the \(s\)-wave grey-body factor vanishes quadratically as \(\omega\to0\). This suppression was shown explicitly for four-dimensional Schwarzschild--de Sitter spacetime in \cite{Crispino:2013pya}. The same matched-expansion structure, with the coefficient fixed by the curvature term across the static patch, also appears in more general asymptotically de Sitter scalar problems, including the dilaton--de Sitter case analyzed in \cite{Ertugay:2026igu}. Consequently, the formal four-dimensional \(s\)-wave cross section has a finite, coupling-dependent low-frequency limit rather than the geometric value \(A_H\). The limit of zero coupling is therefore non-uniform: the minimally coupled case has a finite \(\Gamma_{00}(0)\), whereas any fixed nonzero coupling drives \(\Gamma_{00}\) to zero.

\subsection{High-energy absorption and Regge-pole description}

While the low-frequency absorption cross section is governed by the near-horizon geometry and obeys the universal limit $\sigma_{\mathrm{abs}} \to A_H$, the high-frequency regime is controlled by null geodesics near the photon sphere. In the eikonal limit, $\omega r_H \gg 1$, the scattering problem is therefore well described by geometric optics. Following \cite{Decanini:2010fz}, we study the motion of a massless particle in the equatorial plane of the spherically symmetric black hole (\ref{eq:higherdimensionalmetric}). The cyclic coordinates $t$ and
$\phi$ imply the conserved energy and angular momentum
\begin{equation}
E = f(r)\dot t,
\qquad
L = r^{2}\dot \phi.
\end{equation}
Here a dot denotes differentiation with respect to an affine parameter. The null
condition,
\begin{equation}
2\mathcal{L} = -f(r)\dot t^{2} - \frac{f(r)\dot r^{2}}{F^2(r)} + r^2\dot \phi^2=0,
\end{equation}
can be written as
\begin{equation}\label{eq:radialparticle}
\frac{f^2(r)}{F^2(r)}\dot r^{2} + V_{\mathrm{eff}}(r) = E^{2},
\qquad
V_{\mathrm{eff}}(r) = \frac{L^{2}}{r^{2}}f(r).
\end{equation}
It is convenient to introduce the impact parameter
\begin{equation}
b \equiv \frac{L}{E},
\end{equation}

The circular null orbit at $r=r_c$ is determined by the usual conditions
\begin{equation}
\dot r\big|_{r=r_c}=0,
\qquad
V_{\mathrm{eff}}'(r)\bigg|_{r=r_c}=0.
\end{equation}
These imply
\begin{equation}
b_c^2 = \frac{r_c^2}{f(r_c)},
\qquad
r_c f'(r_c) - 2f(r_c) = 0,
\label{eq:photon-sphere-condition}
\end{equation}
and therefore the critical impact parameter can be written as
\begin{equation}
b_c = \frac{r_c}{\sqrt{f(r_c)}} = \min_{r>r_H}\frac{r}{\sqrt{f(r)}}.
\end{equation}
In particular, $F(r)$ does not affect the photon-sphere condition or the value of $b_c$; it only changes the rate at which the orbit is traversed.

All massless particles with $b<b_c$ are captured by the black hole. For an observer at asymptotic infinity, the shadow radius is therefore equal to the critical impact parameter, and the geometric absorption cross section is the volume of the $(D-2)$-dimensional ball of radius $b_c$,
\begin{equation}
\sigma_{\mathrm{geo}}=\frac{\Omega_{D-3}}{D-2}b_c^{D-2}=\frac{\pi^{(D-2)/2}}{\Gamma\!\left(\frac{D}{2}\right)}b_c^{D-2}.
\end{equation}

To quantify the instability of the circular null orbit, we perturb the radial position as $r(t)=r_c+\delta r(t)$ and expand the radial equation
(\ref{eq:radialparticle}) to second order around $r_c$. One then finds
\begin{equation}
\left(\frac{d}{dt}\delta r\right)^2=\lambda_c^2\delta r^2+\Order{\delta r}^3,
\end{equation}
Since $E^2 = V_{\mathrm{eff}}(r_c)$ and $V_{\mathrm{eff}}'(r_c)=0$, the
coordinate-time Lyapunov exponent is
\begin{equation}
\lambda_c = \sqrt{-F^2(r_c)\frac{V_{\mathrm{eff}}''(r_c)}{2E^2}}.
\end{equation}
Using
\begin{equation}
V_{\mathrm{eff}}''(r_c) = \frac{L^2}{r_c^4}\left[r_c^2 f''(r_c) - 2f(r_c)\right],
\end{equation}
we obtain
\begin{equation}
\lambda_c = \frac{F(r_c)}{r_c}\sqrt{1-\frac{r_c^2 f''(r_c)}{2f(r_c)}}
= \frac{F(r_c)}{r_c}\sqrt{1-\frac{b_c^2 f''(r_c)}{2}}
= \frac{\eta_c}{b_c},
\end{equation}
where we have introduced the following dimensionless quantity
\begin{equation}\label{eq:etadef}
\eta_c \equiv b_c\lambda_c
= \frac{F(r_c)}{\sqrt{f(r_c)}}\sqrt{1-\frac{b_c^2 f''(r_c)}{2}}
= \frac{F(r_c)}{f(r_c)}\sqrt{f(r_c)-\frac{r_c^2}{2}f''(r_c)}.
\end{equation}

We now turn to the high-frequency absorption cross section. When $\omega=k \gg r_H^{-1}$, the higher $\ell$ contributions dominate in the sum (\ref{eq:absorption_spherical}),
\begin{equation}\label{eq:sigma-partial-sum}
\sigma_{\mathrm{abs}}(\omega)
= \frac{\pi^{\frac{D-2}{2}}}{\Gamma\!\left(\frac{D-2}{2}\right)\omega^{D-2}}
\sum_{\ell=0}^{\infty}
\frac{(\ell+D-4)!}{\ell!}(2\ell+D-3)\,\Gamma_\ell(\omega),
\end{equation}
where we used the equality $\Gamma\!\left(\frac{D-2}{2}\right)\Gamma\!\left(\frac{D-1}{2}\right)=\frac{\sqrt{\pi}(D-3)!}{2^{D-2}}$.

It is convenient to introduce the Regge variable
\begin{equation}\label{eq:Regge}
\kappa = \ell + \frac{D-3}{2}.
\end{equation}
Then, the effective potential for the massless scalar field (cf.~Eq.~\ref{eq:effpot_eikonal}) can be written as
\begin{equation}
V_\ell(r) = f(r)\frac{\Lambda_{\ell}}{r^2}
+ \frac{D-2}{2r}F(r)F'(r)
+ \frac{(D-2)(D-4)}{4r^2}F^2(r) = \kappa^2\frac{f(r)}{r^2}\left(1+\Order{\frac{1}{\kappa^{2}}}\right).
\end{equation}
Thus, the top of the potential barrier is located at the photon sphere ($r=r_c$), and the eikonal formula for the grey-body factor (\ref{eq:gb_eikobal}) reads
\begin{equation}\label{eq:greybodyhigh}
\Gamma_\ell(\omega)
\simeq \left[1 + \exp\!\left(-2\pi\,\frac{\omega^2 - V_{\ell}(r_c)}{\sqrt{-2F^2(r_c)V_{\ell}''(r_c)}}\right)\right]^{-1}=\left[1 + \exp\!\left(-\frac{\pi\left(b_c^2\omega^2-\kappa^2\right)}{\eta_c\kappa}\right)\right]^{-1}.
\end{equation}

After analytically continuing the grey-body factor (\ref{eq:greybodyhigh}) to the complex-angular-momentum plane, we apply the half-range Poisson summation formula to Eq.~(\ref{eq:sigma-partial-sum}),
\begin{equation}\label{eq:poisson-half-range}
\sigma_{\mathrm{abs}}(\omega)
= \frac{2\pi^{\frac{D-2}{2}}}{\Gamma\!\left(\frac{D-2}{2}\right)\omega^{D-2}}\sum_{p=-\infty}^{\infty}\int_0^{\infty}\mathcal{F}(\kappa)
\,\exp\left(2\pi\imo p\left[\kappa-\frac{D-3}{2}\right]\right)\,d\kappa,
\end{equation}
where we introduced a complex analytic function,
\begin{equation}
\mathcal{F}(\kappa)
\equiv \frac{\Gamma\!\left(\kappa+\frac{D-3}{2}\right)}{\Gamma\!\left(\kappa-\frac{D-5}{2}\right)}
\,\kappa\,\Gamma_{\kappa-(D-3)/2}(\omega).
\end{equation}

In the limit $\omega\to\infty$,
$$\Gamma_\ell(\omega)\simeq\Theta\!\left(\omega^2-\frac{\kappa^2}{b_c^2}\right), \qquad \frac{\Gamma\!\left(\kappa+\frac{D-3}{2}\right)}{\Gamma\!\left(\kappa-\frac{D-5}{2}\right)}
\simeq \kappa^{D-4},$$
and only $p=0$ contributes to the sum~(\ref{eq:poisson-half-range}). Finally, we obtain
\begin{equation}
\lim_{\omega\to\infty}\sigma_{\mathrm{abs}}(\omega)
= \lim_{\omega\to\infty}\frac{2\pi^{\frac{D-2}{2}}}{\Gamma\!\left(\frac{D-2}{2}\right)\omega^{D-2}}
\int_0^{b_c\omega}\kappa^{D-3}\,d\kappa=\frac{\pi^{(D-2)/2}}{\Gamma\!\left(\frac{D}{2}\right)}b_c^{D-2}=\sigma_{\mathrm{geo}}.
\end{equation}

To obtain the leading correction around $\sigma_{\mathrm{geo}}$, one performs the Watson-Sommerfeld resummation over the residues of the grey-body factors in the complex plane \cite{Decanini:2011xi}. In the eikonal limit, the Regge poles of (\ref{eq:greybodyhigh}) are located at
\begin{equation}
\kappa_n(\omega) = b_c\omega + i\eta_c\left(n-\frac{1}{2}\right) + \Order{b_c\omega}^{-1}, \qquad n = 1,2,\ldots,
\end{equation}
with residues
\begin{equation}
\gamma_n(\omega) = -\frac{\eta_c}{2\pi} + \Order{b_c\omega}^{-1}.
\end{equation}
Keeping only the first Regge pole ($n=1$) then gives the dominant oscillatory part of the absorption cross section,
\begin{equation}\label{eq:sigma-total-eikonal}
\sigma_{\mathrm{abs}}(\omega)
= \sigma_{\mathrm{geo}}\,\left(1+(-1)^{D-3}2(D-2)\eta_c\,e^{-\pi\eta_c}\,
\frac{\sin(2\pi b_c\omega)}{b_c\omega}+\Order{b_c\omega}^{-2}\right).
\end{equation}
This expression shows that the absorption cross section oscillates around its geometric-optics limit with a frequency determined by the orbital period of null geodesics at the photon sphere and an amplitude controlled by the instability of these orbits.

Beyond the leading eikonal approximation, Regge-pole analyses reveal a fine structure in the high-energy absorption cross section \cite{Decanini:2011xw}. The complex-angular-momentum framework has also been applied to the resonance and absorption spectra of massive scalar perturbations of Schwarzschild black holes \cite{Decanini:2011eh}.

Thus, in contrast to the low-energy regime, where the absorption is determined solely by the horizon area, the high-energy behavior is governed by the properties of the photon sphere. The cross section approaches the geometric-optics limit with superimposed oscillations that encode detailed information about the spacetime geometry near the unstable circular orbit. This establishes a complementary picture in which different regions of the spacetime control different frequency regimes: the horizon dominates the infrared behavior, while the photon sphere governs the ultraviolet regime of the absorption spectrum.

\subsection{Frobenius and inverse-scattering methods}\label{sec:MST}

For separable perturbations of Kerr black holes, the horizon-side solution can be expanded in hypergeometric functions, while the solution at infinity can be expanded in Coulomb wave functions \cite{Mano:1996vt}. The coefficients satisfy a three-term recurrence relation, and the minimal-solution condition determines the renormalized angular momentum. Matching the two expansions determines the ingoing, outgoing, and transmitted amplitudes. For separable perturbations of Kerr--Newman--de Sitter black holes, the radial and angular equations can be reduced to Heun equations, for which similar hypergeometric-function expansions can be constructed \cite{Suzuki:1999nn}. The scattering problem is then reduced to solving two nonalgebraic equations with infinite continued fractions.

More generally, many separated black-hole radial equations can be transformed into confluent-Heun form. The grey-body factor can then be formulated directly as a connection problem between local Frobenius bases at the horizon and at infinity, without introducing an intermediate WKB approximation. Monodromy and isomonodromic techniques, together with equivalent representations in terms of irregular conformal blocks, provide relations among the relevant connection coefficients. These relations can be used to calculate grey-body factors, quasinormal modes, and tidal response within a common framework \cite{Bonelli:2021uvf}.

A different inverse-scattering approach applies to real, short-range, one-dimensional effective potentials. In this case, the transmission probability can be related to the conserved integrals of the Korteweg--de Vries hierarchy \cite{Lenzi:2023inn}. These invariants, which are spatial integrals of differential polynomials in the potential, determine moments of the logarithm of the transmission coefficient. A finite set of such moments can be used to reconstruct an approximate grey-body factor through Padé approximation and inverse Laplace transformation. Tests with Schwarzschild effective potentials show that this procedure can recover the transmission curve without integrating a separate scattering solution at every frequency. Extensions to long-range, complex, or frequency-dependent potentials, however, require additional analysis.

\section{The Visser bound on grey-body factors}\label{sec:Visser}

An important non-perturbative result concerning one-dimensional scattering was obtained by Visser in the context of quantum-mechanical potential scattering~\cite{Visser:1998ke}. For perturbation channels whose radial equation is a real one-dimensional Schrödinger problem with well-defined asymptotic plane waves, the same analysis yields rigorous lower bounds on the transmission probability at nonzero asymptotic wavenumber, provided that the relevant integral converges.

\subsection{Visser inequality}

The key idea is to rewrite Eq.~(\ref{schroedinger_GBF}) as a first-order system by introducing an auxiliary real phase function $\varphi(r_*)$ and decomposing the wavefunction as
\begin{equation}
\Psi(r_*) = a(r_*)\,\frac{e^{+i\varphi}}{\sqrt{\varphi'}} + b(r_*)\,\frac{e^{-i\varphi}}{\sqrt{\varphi'}},
\end{equation}
where $\varphi' \equiv d\varphi/dr_* \neq 0$. The functions $a(r_*)$ and $b(r_*)$ play the role of position-dependent Bogoliubov coefficients. Imposing a convenient gauge condition allows the second-order equation to be rewritten as two coupled first-order equations for $a$ and $b$.

When $\varphi$ is chosen to be real, the probability current takes the simple form
\begin{equation}
|a|^2 - |b|^2 = 1,
\end{equation}
which expresses flux conservation. Imposing the standard scattering boundary condition of a purely incoming wave at $r_* \to -\infty$, one sets
\[
a(-\infty)=1, \qquad b(-\infty)=0.
\]
The asymptotic values of the coefficients at $r_* \to +\infty$ define the Bogoliubov amplitudes
\[
\alpha = a(+\infty), \qquad \beta = b(+\infty),
\]
which satisfy $|\alpha|^2 - |\beta|^2 = 1$.
The grey-body factor is then given by
\[
\Gamma_{\ell m}(\omega)=1-|R_{\ell m}|^2=\frac{1}{|\alpha|^2}.
\]

By estimating the evolution of $|a|$ along $r_*$ and bounding the real part of certain complex products, one arrives at the inequality
\begin{equation}
\frac{d|a|}{dr_*}\le\vartheta(r_*)\sqrt{|a|^2 - 1},
\end{equation}
where
\begin{equation}
\vartheta(r_*) = \frac{\sqrt{(\varphi'')^2 + \left[U - (\varphi')^2\right]^2}}{2|\varphi'|}.
\end{equation}
Integration from $r_*=-\infty$ to $r_*=+\infty$ yields
\begin{equation}
|\alpha|\le\cosh\!\left(\int_{-\infty}^{+\infty}\vartheta(r_*)\,dr_*\right),
\end{equation}
and therefore
\begin{equation}
\Gamma_{\ell m}(\omega) = \frac{1}{|\alpha|^2}\ge\left[\cosh\!\left(\int_{-\infty}^{+\infty}\vartheta(r_*)\,dr_*\right)\right]^{-2}.
\label{general_bound}
\end{equation}
This inequality is exact and non-perturbative. It does not rely on WKB validity, smallness of the potential, or any expansion in frequency.

\subsection{Implications for grey-body factors}\label{subsec:visser_implications}

A particularly transparent specialization is obtained by choosing the auxiliary phase to satisfy
\begin{equation}
\varphi'(r_*) = \omega,
\end{equation}
which is admissible when the potential in Eq.~(\ref{eq:schroedinger_simple}) vanishes at both asymptotics, $V_{\ell}(r_*\to\pm\infty)\to 0$.
In this case the bound simplifies drastically:
\begin{equation}
\vartheta(r_*) = \frac{\sqrt{(\varphi'')^2 + \left[\omega^2 - V_{\ell}(r_*) - (\varphi')^2\right]^2}}{2|\varphi'|}=\frac{|V_{\ell}(r_*)|}{2\omega},
\end{equation}
and the grey-body factor~(\ref{general_bound}) satisfies
\begin{equation}
\Gamma_{\ell}(\omega)\ge\left[\cosh\!\left(\frac{1}{2\omega}\int_{-\infty}^{+\infty}|V_{\ell}(r_*)|\,dr_*\right)\right]^{-2}.
\label{GBF_bound_simple}
\end{equation}

Equation~(\ref{GBF_bound_simple}) has several important implications for grey-body factors of black holes.
\begin{enumerate}
\item At any finite frequency, provided the integral of the effective potential is finite, the transmission probability has a strictly positive lower bound. In the low-frequency regime this bound may become very small, but its suppression is controlled explicitly by the integrated strength of the potential barrier.

\item The bound is largely model independent within its assumptions. It applies equally to scalar, electromagnetic, and gravitational perturbations, and to arbitrary black-hole backgrounds whose perturbation equations can be cast in the required one-dimensional scattering form, provided the effective potential decays sufficiently fast at both asymptotic ends. No assumption is made about the existence of turning points.

\item The structure of the inequality makes clear that the total scattering is controlled by the ``area'' of the barrier in the tortoise coordinate. Sharper or higher barriers increase the integral in Eq.~(\ref{GBF_bound_simple}) and thus lower the minimal possible transmission. Conversely, if the potential is smoothed or lowered so that this integral decreases, the lower bound on \(\Gamma_{\ell}(\omega)\) increases. In this sense, narrow or high barriers are more efficient reflectors, whereas smoother barriers are less effective at suppressing transmission.

\item The bound remains valid even in regimes where WKB methods are unreliable. Unlike semiclassical estimates, which may break down near turning points or at low frequency, Eq.~(\ref{GBF_bound_simple}) is a rigorous inequality derived directly from flux conservation and the first-order representation of the wave equation.
\end{enumerate}

Thus, the Visser bound provides a powerful analytic constraint on grey-body factors. While it does not replace detailed numerical calculations, it guarantees that the transmission probability is always bounded from below by a simple integral functional of the effective potential, offering valuable qualitative control over Hawking radiation spectra in a broad class
of black-hole spacetimes \cite{Boonserm:2014rma,Boonserm:2017qcq,Boonserm:2019mon,Chowdhury:2020bdi,Kanzi:2020cyv,Boonserm:2021owk,Sakalli:2022xrb,Javed:2022bdi,Albuquerque:2023lhm,Baruah:2023rhd,Al-Badawi:2023emj,Gogoi:2023fow,Al-Badawi:2023lvx,Jha:2023rem,Al-Badawi:2024iax,Ovgun:2024zmt,Al-Badawi:2024cby,Hosseinifar:2024wwe,Jha:2024ltc,Singh:2024nvx,Rincon:2024won,Sekhmani:2024rpf,Gogoi:2024epx,Nakarachinda:2025hcd,Chaudhary:2025lzi,Sekhmani:2025xzf,Baruah:2025ifh,Media:2025udn,Sajadi:2025kah,Muniz:2025iqu,Singh:2025lrk,Al-Badawi:2025rcq,Javed:2025yea,Ahmed:2025qow,Nakarachinda:2025bvy,Sarkar:2025fjx,Sarkar:2025fao,Sekhmani:2025sml,Sekhmani:2026aun,Belchior:2026nma}.

\section{Numerical methods for computing grey-body factors}\label{sec:numerical}

Numerical methods are needed whenever the effective potential is too complicated for controlled analytic approximations, or when accurate grey-body factors are required over a broad range of frequencies and multipoles.

\subsection{Numerical integration}
The direct numerical calculation of grey-body factors is a connection problem for a second-order differential equation between two singular points. For asymptotically flat and AdS black holes, these points are the event
horizon and spatial infinity, and for de Sitter black holes they are the event horizon and the cosmological horizon. The task is therefore not only to integrate Eq.~(\ref{schroedinger_GBF}), but also to connect the local solution satisfying the physical boundary condition at one singular point to the appropriate pair of independent asymptotic solutions at the other one.

In the standard scattering problem one imposes the transmitted wave at the event horizon,
\begin{equation}\label{eq:numerical_H_expansion}
\Psi_{tran}=e^{-i(\omega-\Phi_H)r_*}\left(1+h_1(r-r_H)+h_2(r-r_H)^2+h_3(r-r_H)^3+\cdots\right).
\end{equation}
The coefficients \(h_1,h_2,h_3,\ldots\) are obtained recursively by substituting this expansion into the radial equation appropriate to the background. The overall normalization is arbitrary; Eq.~(\ref{eq:numerical_H_expansion}) corresponds to setting \(A_{tran}=1\). The numerical integration is then started slightly outside the event horizon, at \(r=r_H+\epsilon_H\), with initial data supplied by the truncated series.

The solution obtained in this way is matched in the asymptotic region to
\begin{equation}\label{eq:numerical_fit}
\Psi(r_*)=A_{in}\Psi_{in}(r_*)+A_{out}\Psi_{out}(r_*),
\end{equation}
where \(\Psi_{in}\) and \(\Psi_{out}\) are the two asymptotic basis functions defined in Sec.~\ref{sec:scattering}.

For fields in asymptotically flat spacetimes, the basis functions have the large-$r$ form
\begin{eqnarray}
\Psi_{in}&=&e^{-ikr_*}\left(1+\frac{a_{1}^{in}}{r}+\frac{a_{2}^{in}}{r^2}+\frac{a_{3}^{in}}{r^3}+\ldots\right),\\
\Psi_{out}&=&e^{+ikr_*}\left(1+\frac{a_{1}^{out}}{r}+\frac{a_{2}^{out}}{r^2}+\frac{a_{3}^{out}}{r^3}+\ldots\right),
\end{eqnarray}
where \(k=\sqrt{\omega^2-\mu^2}\), and \(k=\omega\) for massless fields. The coefficients \(a_n^{in,out}\) are determined by substitution into the wave equation. If the effective potential has a long-range tail, the corresponding logarithmic phase must be included in the exponential before the fit is performed. For AdS black holes the matching is instead to the two power-law behaviors discussed in Eq.~(\ref{asymptoticbehavior}).

The same idea applies in asymptotically de Sitter spacetimes, but the matching is performed near the cosmological horizon \(r_C\). The two basis functions are expanded as
\begin{eqnarray}
\Psi_{in}&=&e^{-i(\omega-\Phi_C)r_*}\left(1+c_1^{in}(r_C-r)+c_2^{in}(r_C-r)^2+c_3^{in}(r_C-r)^3+\ldots\right),\\
\Psi_{out}&=&e^{+i(\omega-\Phi_C)r_*}\left(1+c_1^{out}(r_C-r)+c_2^{out}(r_C-r)^2+c_3^{out}(r_C-r)^3+\ldots\right).
\end{eqnarray}
Thus, after integrating the horizon-normalized solution (\ref{eq:numerical_H_expansion}) toward \(r_C\), one fits Eq.~(\ref{eq:numerical_fit}) in a region where the truncated cosmological-horizon expansion is accurate.

The coefficients \(A_{in}\) and \(A_{out}\) can be extracted by matching the numerically found \(\Psi\) and \(\partial_{r_*}\Psi\) at a single point. In this case,
\begin{equation}\label{eq:numerical_connection}
\left(\begin{array}{c}
A_{in}\\ A_{out}
\end{array}\right)
=
\left(\begin{array}{cc}
\Psi_{in} & \Psi_{out}\\
\partial_{r_*}\Psi_{in} & \partial_{r_*}\Psi_{out}
\end{array}\right)^{-1}
\left(\begin{array}{c}
\Psi\\ \partial_{r_*}\Psi
\end{array}\right),
\end{equation}
where all functions are evaluated at the chosen matching radius.

The grey-body factor then follows from Eq.~(\ref{GBcalc}),
\begin{equation}
\Gamma_{\ell m}(\omega)=1-\left|\frac{A_{out}}{A_{in}}\right|^2.
\end{equation}

Single-point matching is useful as a diagnostic, but by itself it gives little control over the matching error. In practical calculations one therefore fits the numerical solution to Eq.~(\ref{eq:numerical_fit}) over a finite interval. For asymptotically flat matching one may use \(R<r<R+\Delta R\), with \(R\gg r_H\). In de Sitter space the analogous fit is performed in an interval close to, but still inside, the cosmological horizon, for example
\(r_C-\Delta R<r<r_C-\epsilon_C\). The coefficients \(A_{in}\) and \(A_{out}\) are then obtained from an overdetermined least-squares problem on a set of points in the fitting interval \cite{Konoplya:2010vz}. The numerical precision is assessed by repeating the extraction for different truncation orders, starting offsets, and fitting intervals. One increases the number of terms in Eq.~(\ref{eq:numerical_H_expansion}), decreases \(\epsilon_H\), moves the fitting interval farther into the asymptotic region, and increases the number of terms in the asymptotic expansions until \(\Gamma_{\ell m}\) is stable within the desired precision.

Several points are important for numerical stability.
\begin{enumerate}
\item The singular points should be treated analytically. Factoring out the known singular behavior of the solution prevents the integrator from having to resolve the oscillatory singular factor itself.

\item The asymptotic matching error should be monitored independently of the integration error. A fit over several points is usually more robust than matching at a single point, and it also exposes residual dependence on the chosen fitting interval.

\item Low-frequency transmission is often very small. Then \(|A_{out}/A_{in}|\) is close to unity, and the expression \(1-|A_{out}/A_{in}|^2\) may suffer from loss of relative precision. In this regime one can use the equivalent flat-space formula (\ref{eq:greybody}),
\begin{equation}
\Gamma_{\ell m}(\omega)
=\frac{\omega-\Phi_H}{k}\,\left|\frac{A_{tran}}{A_{in}}\right|^2,
\end{equation}
or, in de Sitter space, Eq.~(\ref{eq:greybodydS}),
\begin{equation}
\Gamma_{\ell m}(\omega)
=\frac{\omega-\Phi_H}{\omega-\Phi_C}\,\left|\frac{A_{tran}}{A_{in}}\right|^2.
\end{equation}
These forms avoid subtracting two nearly equal numbers and keep the superradiant sign explicit.
\end{enumerate}

Finally, one should monitor conserved quantities and conditioning. For a real Schrödinger-type equation the Wronskian (\ref{Wronskian}) must be independent of \(r_*\), so that its drift gives a direct measure of accumulated numerical error. The result should also be insensitive to moving the matching interval inside the asymptotic region. Near resonances, for double-barrier potentials, or when one of the local basis functions is
numerically dominant, the two columns of the matrix in Eq.~(\ref{eq:numerical_connection}) can become nearly linearly dependent or very different in norm. The resulting large condition number amplifies small errors in \(\Psi\), \(\partial_{r_*}\Psi\), and the asymptotic series coefficients. In such cases it is preferable to integrate two independent solutions from the event horizon and from the asymptotic side and match their Wronskians in an intermediate region, or to use a stabilized transfer/scattering-matrix or logarithmic-derivative formulation with rescaling \cite{Johnson:1973ue,Salazar:2026sxp}. Where applicable, comparison with the WKB formula (see Sec.~\ref{sec:WKB}) provides an additional check; however, direct integration remains the method of choice when the effective potential is not a single smooth barrier.

\subsection{Public codes for grey-body factors}

A simplified version of the direct numerical calculation is implemented in the public Mathematica{\textregistered} package \texttt{GrayHawk} \cite{Calza:2025whq}. The code computes grey-body factors from transmission coefficients for massless fields, for which \(k=\omega\), in static, spherically symmetric, asymptotically flat black-hole backgrounds. In this class of geometries, the azimuthal degeneracy is not lifted. The calculation can therefore be performed for fixed \(\omega\) and \(\ell\), with the angular separation constant determined by the spin and angular momentum of the perturbing field. The package provides effective potentials for fields with spins \(s=0,1/2,1,2\), and its modular structure allows the user to specify a new static, spherically symmetric metric and generate numerical values of \(\Gamma_{\ell}(\omega)\). The recently developed \texttt{GrayHawk v2} allows one to work with a numerically computed tortoise coordinate and extends the method to spherically symmetric traversable wormholes \cite{Calza:2026wuf}.

The matching strategy used in \texttt{GrayHawk} is deliberately economical. The numerical solution is fitted, at a discrete set of radial points whose coordinate values are chosen in the implementation to be integers, to a superposition of ingoing and outgoing exponentials, rather than to the full asymptotic expansion in Eq.~(\ref{eq:numerical_fit}), with the coefficients \(a_n^{\rm in,out}\) determined order by order. This makes the code convenient for scans over different metrics and frequencies. At the same time, the accuracy of the extracted amplitudes depends on the distance of the fitting region from the black hole and on the size of the subleading asymptotic corrections. For precision applications, especially at low frequencies or for slowly decaying effective potentials, the output should therefore be checked by varying the fitting region, increasing the numerical precision, and comparing with known analytic limits.

A similar numerical approach has recently been developed for a massive scalar field in the Schwarzschild background. The corresponding public code, shared in Ref.~\cite{Cui:2024uwk}, can in principle be modified to study other backgrounds. In addition, a standalone \texttt{C++} code for asymptotically de Sitter black holes in massive gravity was presented in Ref.~\cite{Dong:2015qpa}. However, that implementation is tailored to the specific model considered there and would require substantial adaptation before it could be used to compute grey-body factors for other black-hole spacetimes.

Another widely used public tool, especially in primordial-black-hole applications, is \texttt{BlackHawk} \cite{Arbey:2019mbc,Arbey:2021mbl}. Its main \texttt{C} routines are designed to compute Hawking evaporation spectra for individual black holes or black-hole distributions, including primary spectra, secondary spectra after hadronization and decays, and the time evolution of the black-hole mass and spin. The grey-body factors enter as precomputed or internally tabulated transmission probabilities that are summed over particle species and angular modes. In this sense, \texttt{BlackHawk} is primarily a phenomenological Hawking-spectrum code \cite{Cheek:2021odj}, while the grey-body-factor calculation is part of the numerical infrastructure that supplies its emission tables.

The grey-body factors used in \texttt{BlackHawk} have been obtained by numerical integration and asymptotic matching, using a Schrödinger-like equation written in terms of a tortoise coordinate and a single-point extraction of the amplitudes analogous to Eq.~(\ref{eq:numerical_connection}). This approach is efficient and works well in many static cases, and it has been used to produce tables for Schwarzschild, Kerr, and selected extensions or non-standard spherically symmetric backgrounds. However, recent work has shown that the implementation based on the Chandrasekhar tortoise coordinate is not sufficiently stable for near-extremal Kerr black holes in the superradiant regime \cite{Arbey:2025dnc}. In that regime, the coordinate transformation can introduce spurious singular behavior, and the resulting grey-body factors may overestimate the superradiant amplification. The improved treatment avoids this instability by working directly with a rescaled radial coordinate. This method reproduces the known near-horizon and asymptotic behavior more accurately and is implemented in the recently published new version of the package \cite{Arbey:2026koc}.

\subsection{Symbolic-regression surrogates}

A different numerical strategy is to use accurate grey-body factors, obtained from direct integration, as training data for fast analytic surrogate models. This approach was developed in Ref.~\cite{Yuan:2025eyi}, where the \texttt{ReGrayssion} pipeline was introduced to construct compact symbolic-regression approximations for Schwarzschild grey-body factors. The aim is not to replace the first-principles scattering calculation, but to compress a numerical table into simple formulae that can be evaluated rapidly in phenomenological applications, especially in primordial-black-hole evaporation studies where the Hawking spectrum must be computed many times. The input data were generated with \texttt{GrayHawk} for Schwarzschild black holes. The symbolic-regression step was performed with \texttt{PySR}, which searches over analytic expressions built from elementary operators and ranks candidate formulae by a combined measure of loss and expression complexity. To reduce overfitting and test the robustness of the symbolic expressions, the training data were perturbed by Gaussian noise. The analysis found stable results for a noise amplitude of about \(1\%\), which was then used in the final fits.

The symbolic-regression search suggested a simple universal structure for the spin-dependent grey-body factors,
\begin{equation}\label{eq:regression_sr}
\Gamma_{\ell}(\omega)=\operatorname{erf}\!\left[\operatorname{erfc}\!\left(a_s+b_s\ell+c_s\omega\right)\right],
\end{equation}
where the coefficients \(a_s,b_s,c_s\) are fitted separately for each spin and the frequency variable is taken in the normalization used for the training data. In addition, a more transparent ``human regression'' ansatz was considered,
\begin{equation}\label{eq:regression_hr}
\Gamma_{\ell}(\omega)=\frac{1}{2}\left[1+\operatorname{erf}\!\left(a_s+b_s\ell+c_s\omega\right)\right].
\end{equation}
Both formulae encode the expected sigmoid-like transition from suppressed transmission below the effective barrier to almost complete transmission at high frequency. The comparison with the numerical \texttt{GrayHawk} data showed that the formulae reproduce the main behavior of the Schwarzschild grey-body factors for the relevant low multipoles, with residuals typically at the level needed for the evaporation applications considered there. To avoid using the surrogate outside its reliable domain, the analysis imposed a cutoff when the relative residual exceeded a prescribed threshold.

As an application, the analytic grey-body-factor formulae were inserted into the Hawking photon spectrum of a monochromatic population of Schwarzschild primordial black holes. The resulting extragalactic $\gamma$-ray flux was compared with observations of the diffuse background. The bounds on the primordial-black-hole fraction obtained with the symbolic-regression and human-regression surrogates were found to be consistent with those obtained from the direct numerical \texttt{GrayHawk} data.

The method is therefore best understood as a data-driven compression of a trusted numerical calculation. Its strengths are speed, interpretability, and ease of use in parameter scans. Its limitations are equally important: the published fits are calibrated for Schwarzschild grey-body factors and for the range of spins, multipoles, and frequencies used in the training set. They should not be extrapolated blindly to rotating, charged, higher-dimensional, or modified-gravity black holes, nor to frequency ranges where the residuals were not tested. For such cases, the surrogate construction would have to be repeated with new numerical data and validated against flux conservation, known asymptotic limits, and, where available, independent calculations.

\subsection{Spectral collocation with radiation functions}

As an alternative to direct numerical integration of the radial wave equation, followed by fitting the resulting solution at large \(r_*\), one may discretize the scattering problem from the outset. This strategy was proposed in Ref.~\cite{Clemente:2014}, where the problem is treated by spectral collocation: a finite-difference-like method in which the differential equation is replaced by a system of algebraic equations on a discrete grid. One constructs two auxiliary real solutions whose behavior is fixed asymptotically,
\begin{equation}\label{eq:spectral_CS_inf}
C(r_*)\to \cos(\omega r_*),\quad S(r_*)\to \sin(\omega r_*),\qquad r_*\to+\infty .
\end{equation}
To satisfy the asymptotic boundary condition
\begin{equation}
\Psi(r_*)\to e^{-i\omega r_*}+R_{\ell m}e^{+i\omega r_*},\qquad r_*\to+\infty,
\end{equation}
which is the spatial-infinity part of Eq.~(\ref{eq:scattering}), the standard unit-incident solution is written as
\begin{equation}\label{eq:spectral_CS_solution}
\Psi(r_*)=(1+R_{\ell m})C(r_*)+i(R_{\ell m}-1)S(r_*).
\end{equation}

The scattering coefficients are then fixed by the behavior of \(C(r_*)\) and \(S(r_*)\) at the event horizon. In the notation of this review,
\begin{equation}\label{eq:spectral_H_coeffs}
\begin{array}{rcl}
C(r_*)&\to&
\alpha_1\cos\left(\omega r_*\right)
+\alpha_2\sin\left(\omega r_*\right),\\
S(r_*)&\to&
\beta_1\cos\left(\omega r_*\right)
+\beta_2\sin\left(\omega r_*\right),
\end{array}
\qquad r_*\to-\infty.
\end{equation}
Matching Eq.~(\ref{eq:spectral_CS_solution}) to the transmitted wave
\(T_{\ell m}e^{-i\omega r_*}\) at the event horizon gives
\begin{equation}\label{eq:spectral_RT_system}
\begin{array}{rcl}
(1+R_{\ell m})\alpha_1+i(R_{\ell m}-1)\beta_1&=&T_{\ell m},\\
(1+R_{\ell m})\alpha_2+i(R_{\ell m}-1)\beta_2&=&-iT_{\ell m}.
\end{array}
\end{equation}
Thus the spectral part of the calculation reduces to finding the four real coefficients
\(\alpha_1,\alpha_2,\beta_1,\beta_2\), which are determined by expanding only the deviations from the prescribed sine and cosine waves at spatial infinity. The auxiliary solutions are represented at spectral truncation order $N\geq2$ as
\begin{equation}\label{eq:spectral_expansion}
\begin{array}{rcl}
C(r_*)&=&\displaystyle{\cos(\omega r_*)+\sum_{n=0}^{N-2}a_n\varphi_n(r_*)+a_{N-1}{\cal C}_H(r_*)+a_N{\cal S}_H(r_*)},\\
S(r_*)&=&\displaystyle{\sin(\omega r_*)+\sum_{n=0}^{N-2}b_n\varphi_n(r_*)+b_{N-1}{\cal C}_H(r_*)+b_N{\cal S}_H(r_*)}.
\end{array}
\end{equation}
Thus, each auxiliary solution contains $N+1$ unknown expansion coefficients, $a_0,\ldots,a_N$ or $b_0,\ldots,b_N$, which are determined from the same number of collocation equations. The basis functions are recombined rational Chebyshev functions,
\begin{equation}
\varphi_n(r_*)=\frac{1}{2}\left[T_{n+2}(y(r_*))-T_n(y(r_*))\right],
\qquad n=0,\ldots,N-2,
\end{equation}
where \(T_n(y)\) is the ordinary Chebyshev polynomial of the compact coordinate, which is fixed via
\begin{equation}\label{eq:rational_chebyshev_map}
\frac{r_*-r_*^{(0)}}{L_0}=\frac{y}{\sqrt{1-y^2}},\qquad -1<y<1 .
\end{equation}
Here $L_0>0$ is the mapping scale, with the same dimensions as $r_*$, and controls the distribution of collocation points over the physical domain. The parameter $r_*^{(0)}$ is the center of the map: $y=0$ corresponds to $r_*=r_*^{(0)}$. It is conveniently chosen near the region where the effective potential is appreciable.

Because \(T_{n+2}\) and \(T_n\) have the same limiting values at \(y=\pm1\), the functions \(\varphi_n(r_*)\) vanish as \(r_*\to\pm\infty\). The last two functions in Eq.~(\ref{eq:spectral_expansion}) carry the event-horizon sine and cosine behavior while vanishing at spatial infinity. A convenient choice is
\begin{eqnarray}
{\cal C}_H(r_*)&=&\frac{1-\tanh\left[\epsilon_0(r_*-r_*^{(0)})\right]}{2}\cos\left[\omega r_*\right],\\
{\cal S}_H(r_*)&=&\frac{1-\tanh\left[\epsilon_0(r_*-r_*^{(0)})\right]}{2}\sin\left[\omega r_*\right].
\end{eqnarray}
The parameter $\epsilon_0>0$ has dimensions of inverse length and controls how rapidly these radiation functions switch from their horizon behavior to zero. The transition is centered at $r_*^{(0)}$ and has a characteristic width of order $\epsilon_0^{-1}$. Since \({\cal C}_H\to\cos(\omega r_*)\) and \({\cal S}_H\to\sin(\omega r_*)\) at the event horizon, the asymptotic coefficients are read directly from the last two spectral coefficients,
\begin{equation}
\alpha_1=1+a_{N-1},\qquad \alpha_2=a_N,\qquad
\beta_1=b_{N-1},\qquad \beta_2=1+b_N .
\end{equation}

The coefficients \(a_n\) and \(b_n\) are obtained by collocation. One substitutes Eq.~(\ref{eq:spectral_expansion}) into Eq.~(\ref{schroedinger_GBF}) and sets the residuals to zero at the rational Chebyshev points
\begin{equation}
y_j=\cos\frac{(2j+1)\pi}{2N+2},\qquad j=0,\ldots,N\,.
\end{equation}
This gives two linear algebraic systems, one for the \(a_n\) and one for the \(b_n\). After these systems are solved, Eq.~(\ref{eq:spectral_RT_system}) yields \(R_{\ell m}\) and \(T_{\ell m}\), and the grey-body factor follows from Eq.~(\ref{eq:greybody}),
\begin{equation}\label{eq:flux-identity}
\Gamma_{\ell m}(\omega)=1-|R_{\ell m}|^2=|T_{\ell m}|^2.
\end{equation}

The accuracy of the method is controlled by increasing \(N\), varying the map parameters \(L_0\) and \(r_*^{(0)}\), and changing the transition scale \(\epsilon_0\). The recombined functions \(\varphi_n\) impose the endpoint decay analytically, leaving the last two basis functions to carry the radiation data. In practice, one should also monitor the residuals of the collocation equations and the flux-conservation identity~(\ref{eq:flux-identity}), because the dense spectral matrices can become ill-conditioned when \(N\) is too large or when the map does not resolve the region where the effective potential is appreciable.

\section{Correspondence between quasinormal modes and grey-body factors}\label{sec:correspondence}

Quasinormal modes of asymptotically flat black holes are characteristic complex frequencies associated with the ringdown stage of black-hole perturbations. They correspond to solutions of the master wave equation obeying purely ingoing-wave boundary conditions at the event horizon and purely outgoing-wave boundary conditions at infinity. Mathematically, quasinormal frequencies can be interpreted as resonances, or poles of the scattering matrix in the complex frequency plane. They are therefore determined by the same effective potential that governs grey-body factors, albeit under different boundary conditions \cite{Kokkotas:1999bd,Nollert:1999ji,Konoplya:2011qq,Bolokhov:2025rng}.

An indication of a direct link between grey-body factors and quasinormal modes appeared in studies of the gravitational-wave spectral amplitude during the ringdown phase. It was observed that for the dominant mode $(\ell,m)=(2,2)$ the frequency-domain strain measured at infinity acquires an imprint of the grey-body factor, which is discussed in Sec.~\ref{sec:ringdown_gbf}. In particular, near the ringdown frequency $f_{\ell m}\equiv\re{\omega_{\ell m 0}}$, the spectral amplitude can be written as \cite{Oshita:2023cjz}
\begin{equation}
|\tilde h_{\ell m}(\omega)|\simeq c_{\ell m}\,\gamma_{\ell m}(\omega)=c_{\ell m}\,\frac{\sqrt{1-\Gamma_{\ell m}(\omega)}}{\omega^3},
\qquad\omega \gtrsim f_{\ell m},
\label{h_GBF_first}
\end{equation}
where $\Gamma_{\ell m}(\omega)$ is the grey-body (transmission) factor and $c_{\ell m}$ is a constant determined by the source amplitude. Equation~(\ref{h_GBF_first}) shows that the observed gravitational-wave spectrum is not linked solely to the quasinormal pole structure, but is modulated by the frequency-dependent transmission probability through the effective potential barrier.

In \cite{Konoplya:2024lir}, the correspondence between quasinormal modes and grey-body factors was established using WKB calculations of both quantities. When the effective potential $V(r_*)$ entering the wave equation~(\ref{schroedinger_GBF}) forms a single-peaked barrier, the WKB approximation provides a convenient tool for determining the dominant quasinormal frequencies. These modes satisfy the boundary conditions
\begin{equation}\label{boundaryconditions}
\Psi(r_*\to\pm\infty)\propto e^{\pm\imo \omega r_*},
\end{equation}
which correspond to purely ingoing waves at the event horizon ($r_*\to -\infty$) and purely outgoing waves at spatial infinity ($r_*\to +\infty$).
The resulting eigenvalue problem can be solved using the WKB method. In particular, Eq.~(\ref{eq:WKBexpansion}) may be employed to determine the complex quasinormal frequencies, labeled by
\begin{equation}\label{eq:qnms_quantization}
\K = n + \frac{1}{2}, \qquad n=0,1,2,\ldots,
\end{equation}
where $n$ labels the overtone number. We denote the corresponding complex frequencies by $\omega_0,\omega_1,\omega_2,\ldots$ in order to distinguish them from the real wave frequency $\omega$. Here $\omega_0$ represents the fundamental mode, which has the slowest decay rate, while $\omega_1$ denotes the first overtone, and so forth. Eq.~(\ref{eq:gb_wkb}) further implies that the quantization conditions~(\ref{eq:qnms_quantization}) identify the quasinormal modes as poles of the analytic continuation of the transmission amplitude into the complex $\omega$-plane.

Since the WKB approach yields corrections to the eikonal approximation for both grey-body factors and quasinormal modes, these corrections can be systematically eliminated from the grey-body factor expression in favor of the quasinormal frequencies, order by order. In the spherically symmetric case, this is achieved by considering the high-multipole expansion of the effective potential $V_{\ell}(r_*)$ in terms of the Regge variable~(\ref{eq:Regge}), $\kappa\sim\ell$,
\begin{equation}\label{potential-multipole}
V_{\ell}(r_*)=\kappa^2\V_0(r_*)+\kappa \V_1(r_*)+\V_2(r_*)+\kappa^{-1}\V_3(r_*)+\ldots,
\end{equation}
where successive terms contribute at progressively higher orders in the WKB expansion.

In particular, the eikonal approximation~(\ref{eq:gb_eikobal}) depends on the value of $\V_0$ and its second derivative evaluated at the maximum of the effective potential,
\begin{equation}\label{eq:WKBformula_eikonal}
\omega = \sqrt{V_{\ell} - \imo \K \sqrt{-2V_{\ell}''}} + \Order{\kappa^{-1}} = \kappa\sqrt{\V_0} - \imo \K \sqrt{\frac{-\V_0''}{2\V_0}} + \Order{\kappa^{-1}}.
\end{equation}
At the same time, Eq.~(\ref{eq:qnms_quantization}) implies that the dominant quasinormal frequency ($n=0$, $\K=1/2$) is given by
\begin{eqnarray}\label{eq:WKBformula_eikonal_dominant}
\omega_0 &=& \kappa\sqrt{\V_0} - \frac{\imo}{2}\sqrt{\frac{-\V_0''}{2\V_0}} + \Order{\kappa^{-1}}.
\end{eqnarray}
Eliminating $\V_0$ and $\V_0''$ from Eqs.~(\ref{eq:WKBformula_eikonal}) and (\ref{eq:WKBformula_eikonal_dominant}), and substituting $\K$ into (\ref{eq:gb_wkb}), we obtain
\begin{eqnarray}\label{eq:transmission-eikonal}
\Gamma_{\ell m}(\omega)\equiv |T|^2 &=& \left(1+e^{2\pi\dfrac{\omega^2-\re{\omega_0}^2}{4\re{\omega_0}\im{\omega_0}}}\right)^{-1} + \Order{\frac{1}{\kappa}} .
\end{eqnarray}

This expression (\ref{eq:transmission-eikonal}) links the transmission probability $\Gamma_{\ell}(\omega)$ directly to the lowest quasinormal frequency $\omega_0$ of a spherically symmetric black hole. In the eikonal limit this relation becomes exact, whereas for moderate or small multipole numbers it should be regarded as an approximation.
Its accuracy can nevertheless be improved beyond the strict eikonal regime by incorporating higher-order WKB corrections, expressed in terms of higher overtones. At the third WKB order, the grey-body factors are obtained from Eq.~(\ref{eq:WKBformula_eikonal}), where the parameter $\K$ is expressed through the real frequency $\omega$ together with the fundamental mode $\omega_0$ and the first overtone $\omega_1$ \cite{Konoplya:2024vuj,Konoplya:2024lir},
\begin{eqnarray}\nonumber
i\K&=&\frac{\omega^2-\re{\omega_0}^2}{4\re{\omega_0}\im{\omega_0}}
\left(1+\frac{(\re{\omega_0}-\re{\omega_1})^2}{32\im{\omega_0}^2}
-\frac{3\im{\omega_0}-\im{\omega_1}}{24\im{\omega_0}}\right)
-\frac{\re{\omega_0}-\re{\omega_1}}{16\im{\omega_0}}
\\\nonumber
&&-\frac{(\omega^2-\re{\omega_0}^2)^2}{16\re{\omega_0}^3\im{\omega_0}}
\left(1+\frac{\re{\omega_0}(\re{\omega_0}-\re{\omega_1})}{4\im{\omega_0}^2}\right)
\\\nonumber
&&+\frac{(\omega^2-\re{\omega_0}^2)^3}{32\re{\omega_0}^5\im{\omega_0}}
\left(1+\frac{\re{\omega_0}(\re{\omega_0}-\re{\omega_1})}{4\im{\omega_0}^2}\right.
\\
&&\left.\qquad+\re{\omega_0}^2
\left(\frac{(\re{\omega_0}-\re{\omega_1})^2}{16\im{\omega_0}^4}
-\frac{3\im{\omega_0}-\im{\omega_1}}{12\im{\omega_0}}\right)\right)
+\Order{\frac{1}{\kappa^3}} .
\label{eq:gbsecondorder}
\end{eqnarray}

In numerous recent studies, this correspondence beyond the eikonal limit has been extensively tested \cite{Malik:2024cgb,Skvortsova:2024msa,Han:2025cal,Lutfuoglu:2025hjy,Lutfuoglu:2025eik,Lutfuoglu:2025ljm,Huang:2025rxx,Antoniou:2026jvh,Han:2026fpn}. These tests show that, typically, for $\ell>s$, where $s$ denotes the spin of the field, the relation holds with high accuracy, with the relative error usually remaining well below one percent.

At the same time, when the effective potential has a more complicated structure -- for example, a double-well form -- the correspondence may fail or exhibit significant errors beyond the eikonal regime (see the discussion in \cite{Konoplya:2025hgp}).

The breakdown is related to the fact that the WKB approach, which serves as the common basis for determining both grey-body factors and quasinormal modes, ceases to be reliable. Several typical situations where this occurs can be identified:

\begin{enumerate}

\item The WKB method may fail to reproduce the entire spectrum even at large multipole numbers, describing only a part of it. This occurs, for example, for asymptotically de Sitter black holes \cite{Konoplya:2022gjp,Konoplya:2025mvj}. In this case the correspondence remains valid only for the portion of the spectrum captured by the WKB approximation. Consequently, $\omega_0$ and $\omega_1$ should not necessarily be interpreted as the fundamental mode and the first overtone of the full spectrum, but rather as the two lowest modes of the Schwarzschild branch. In addition, there exists another, nonperturbative branch corresponding to de Sitter modes modified by the presence of the black hole \cite{Lopez-Ortega:2012xvr,Lopez-Ortega:2007vlo}.

\item The centrifugal term may deviate from the standard form (\ref{eq:effpot_eikonal}) or even develop a double-well structure in the eikonal regime. As discussed in Sec.~\ref{sec:EikonalWKB}, such situations arise in various theories involving higher-curvature corrections or nontrivial matter content. In these cases, the correspondence between the eikonal quasinormal modes and the properties of null geodesics can be partially or completely violated.
A notable example is provided by asymptotically de Sitter black holes, where an additional, de Sitter, branch of modes appears, which already exists in empty de Sitter space \cite{Lopez-Ortega:2006tjo,Lopez-Ortega:2006aal,Malik:2026oxx}. This branch cannot be captured within the WKB framework, and therefore the correspondence is valid only for the Schwarzschild-type branch of modes \cite{Konoplya:2022gjp}. In all such cases the well-known correspondence between quasinormal modes and null geodesics \cite{Cardoso:2008bp} is violated \cite{Konoplya:2017wot,Konoplya:2020bxa,Konoplya:2022gjp}, and consequently the relation between quasinormal modes and grey-body factors also breaks down.

\item If the boundary conditions are different, as in anti-de Sitter spacetime, the standard WKB method cannot be applied, and consequently the correspondence cannot be established \cite{Konoplya:2019hlu}.

\item Finally, the correspondence may become inaccurate when the matching between the Taylor expansion of the potential near its maximum and the WKB series is poorly controlled. In such situations the relative error of the approximation may become significant.

\end{enumerate}

For rotating black holes the correspondence also fails in the regime of superradiant scattering, as demonstrated in \cite{Konoplya:2024vuj}.

\section{Scattering phenomenology of compact objects}\label{sec:blackholes}

The preceding sections developed general methods for defining and computing grey-body factors. We now turn to their behavior in concrete black-hole and compact-object spacetimes. The purpose of this section is not to give an exhaustive catalogue of all known calculations, but to highlight representative situations in which grey-body factors probe the structure of the effective potential, distinguish different geometries, or provide observables complementary to quasinormal modes and Hawking spectra.

\subsection{Stability of grey-body factors in astrophysical environments}

The ringdown applications discussed in Sec.~\ref{sec:ringdown_gbf} motivate the use of grey-body factors as observables of the remnant geometry. A grey-body factor, however, is a real-frequency scattering quantity for a specified wave equation and background, whereas a quasinormal frequency is a pole obtained after analytic continuation of the same scattering problem. Consequently, the strong spectral instability of some quasinormal modes \cite{Nollert:1996rf,Jaramillo:2020tuu,Jaramillo:2021tmt,Destounis:2021lum} does not by itself imply a comparable instability of the transmission probability. Conversely, the stability of the latter under a chosen class of potential deformations does not imply that radiation reaches a distant observer without environmental contamination.

It is useful to distinguish two different possibilities. In the first, surrounding matter participates in the scattering problem itself. It can backreact on the metric and thereby change the effective potential, or it can couple directly to the perturbing field and alter its propagation equation in the vicinity of the black hole. The resulting transmission probability is then the grey-body factor of the full, ``dirty'' configuration, rather than that of the corresponding vacuum black hole. In the second possibility, a wave that has already crossed the principal curvature potential is absorbed, scattered, or reprocessed on its way to the detector. Such post-barrier propagation is not part of the intrinsic grey-body factor unless the intervening medium is included explicitly in the scattering problem. This distinction is especially important for electromagnetic radiation and Hawking particles. For gravitational-wave ringdown, the relevant grey-body factor is instead that of the gravitational perturbation equation: electromagnetic opacity does not directly attenuate gravitational waves, although the stress-energy of the environment and couplings to other perturbation sectors can modify their generation and propagation.

The first possibility has been studied in explicit dirty-black-hole geometries. For a Schwarzschild black hole surrounded by a thin spherical shell, the shell modifies the effective potential and imposes a matching condition on the radial field, so the calculated transmission coefficient already refers to the complete configuration \cite{Macedo:2015ikq}. The low-frequency absorption cross section remains equal to the area of the inner event horizon, while the high-frequency limit is governed by the null-geodesic capture cross section of the full spacetime. At intermediate frequencies, the shell changes the magnitude and oscillatory structure of the absorption spectrum and can, for suitable parameters, generate a second light ring and an intervening potential well. Thus, surrounding matter does not erase the grey-body spectrum but replaces the vacuum result with the spectrum of a different scattering system.

A self-consistent model of a black hole embedded in a galactic halo leads to the same general conclusion \cite{Cardoso:2021wlq}. The leading environmental correction to the ringdown is a gravitational redshift, although the galactic potential also affects wave propagation. For electromagnetic perturbations, making the halo more compact or increasing its mass lowers the effective barrier and increases the transmission probability \cite{Konoplya:2021ube}. However, for parameters appropriate to observed galaxies, the characteristic halo scale is much larger than the black-hole mass, and the deviations of both the quasinormal frequencies and grey-body factors from their Schwarzschild values are insignificant. Large effects found for very compact halos therefore characterize strongly modified configurations rather than realistic environmental contamination.

The practical size of such corrections was examined systematically by introducing localized deformations either near the event horizon or in the far zone. Far-zone deformations, which model structures such as accretion disks, have a negligible effect unless the additional barrier becomes comparable in height to the principal black-hole potential \cite{Konoplya:2025ixm}. Realistic environments are both too dilute and too extended to produce such a barrier. Even deliberately exaggerated far-zone deformations lead only to very small changes in the total emissivity. Their effect is reduced further because the induced change in the grey-body factor oscillates as a function of frequency, causing substantial cancellations after frequency integration. Near-horizon deformations produce larger relative changes, but these changes are concentrated mainly at low frequencies, where the undeformed grey-body factor is already close to zero and the absolute transmitted flux is strongly suppressed \cite{Konoplya:2025ixm}. Consequently, a large percentage difference in this regime need not correspond to a significant change in an observable spectrum. Thus, the deformations expected from realistic environments are not practically relevant to the grey-body spectrum, while substantial effects would instead indicate strong near-horizon modifications rather than ordinary matter surrounding the black hole.

This interpretation is consistent with controlled stability tests using localized bumps in the effective potential. For a smooth Pöschl--Teller deformation, pointwise and frequency-integrated changes in $\Gamma_{\ell m}(\omega)$ remain of the order of the perturbation amplitude even when individual quasinormal frequencies are strongly displaced \cite{Rosato:2024arw}. The modulus of the scattering amplitude is stable in this sense, although its phase, and therefore its real and imaginary parts separately, can remain sensitive to a distant deformation \cite{Kyutoku:2022gbr}. A self-consistent effective-field-theory construction likewise finds that the grey-body factor is stable in the frequency interval relevant to the principal ringdown signal \cite{Oshita:2024fzf}. Sharp secondary barriers can affect the exponentially small high-frequency reflectivity and produce resonant oscillations, but the associated change in the time-domain waveform remains small. Related analyses using equal-amplitude or equal-energy perturbations, piecewise-parabolic barriers, and realistic environmental waveforms support the same conclusion while retaining these frequency-dependent qualifications \cite{Wu:2024ldo,Xie:2025jbr,Spieksma:2024voy}.

The second possibility concerns contamination after the wave has crossed the principal curvature barrier. It is illustrated by coupled electromagnetic and gravitational perturbations in a plasma around a charged black hole \cite{Cannizzaro:2024yee}. If the plasma reaches the light-ring region, electromagnetic-led modes can be screened and, through coupling between the electromagnetic and gravitational sectors, the gravitational ringdown can also be shifted. If a low-density cavity instead separates the black hole from an outer plasma barrier, outgoing electromagnetic waves can be reflected back toward the black hole and generate gravitational-wave echoes. This does not contradict the stability of the intrinsic grey-body factor under weak geometric deformations: it describes an additional propagation effect that must be included in the transfer from the strong-field region to the observer.

The two mechanisms therefore require different treatments. Weak environmental changes to the geometry or effective potential generally leave the grey-body factors practically unchanged in the frequency range that contributes appreciably to the observed signal. By contrast, absorption or reprocessing outside the principal scattering region can modify the measured electromagnetic or particle spectrum even when the intrinsic grey-body factor is stable. Such effects require a separate propagation model unless the entire medium is incorporated into the scattering calculation. For gravitational-wave observations, direct electromagnetic opacity is irrelevant, but environmental stress-energy and couplings to other perturbation sectors may still have to be modeled. Grey-body factors should consequently be regarded as robust complementary observables rather than as quantities wholly immune to the astrophysical environment.

\subsection{Model discrimination and observational degeneracies}

The stability discussed above addresses whether a weak environmental deformation destroys the scattering signal, but it does not establish that the signal identifies a unique black-hole model. A grey-body factor belongs to a specified perturbation channel on a specified background and is determined by the corresponding effective potential and boundary conditions, rather than directly by the name of the underlying gravitational theory. If two theories predict the same exterior geometry, linearized perturbation equation, and boundary conditions for the observed channel, their grey-body factors are identical in that channel. Conversely, the same background metric can lead to different transmission probabilities when the field content, nonminimal couplings, or perturbation equations differ. Model discrimination must therefore be formulated at the level of the complete scattering problem, including the spin and quantum numbers of the field, rather than through a comparison of background metrics alone.

There is also an intrinsic inverse-problem degeneracy. The grey-body factor contains the modulus of a transmission or reflection amplitude but not its scattering phase. Distinct effective potentials can have the same, or nearly the same, transmission probability. Darboux-related potentials and the associated Korteweg--de Vries invariants provide explicit examples of scattering information shared by different potential representations \cite{Lenzi:2023inn}. Restricting the measurement to a finite frequency interval makes the nonuniqueness still greater. The full complex scattering amplitudes contain additional information, but in an observed waveform their phases are combined with the phase of the source and with propagation effects. Moreover, when perturbation sectors are coupled, as can occur in modified gravity or in the presence of matter, the relevant object is a scattering matrix containing both transmission and mode-conversion probabilities. A difference between two theoretical grey-body curves thus demonstrates sensitivity to the models, but does not necessarily mean that the models can be distinguished observationally.

The discriminating information is not distributed uniformly over frequency. In the low-frequency limit, universal results such as the equality between the absorption cross section of a minimally coupled massless scalar and the horizon area reduce sensitivity to many details of the geometry (see Sec.~\ref{sec:lowfrequencymatching}). At high frequencies, absorption is governed mainly by the classical capture region and the photon sphere, so models with similar critical impact parameters can again be difficult to separate (see Sec.~\ref{sec:EikonalWKB}). Caution is therefore required when a grey-body spectrum is inferred from a shadow or from a small number of eikonal quasinormal modes \cite{Pedrotti:2025idg}. For a single-barrier problem, the most useful model-dependent information usually lies in the intermediate-frequency transition between reflection and transmission, where the position, height, and curvature of the barrier determine the location and width of the transition. Additional distinctive features can arise from the boundaries of a superradiant interval, field-mass thresholds, resonant transmission through multiple barriers, or the relative behavior of different multipoles and polarizations.

A meaningful comparison must first place the models on the same physical footing. Their asymptotic masses and angular momenta should be matched, the same perturbing field and angular channel should be used, and frequencies should be expressed in a common dimensionless normalization, such as $M\omega$. Even then, $M$ and the rotation parameter are inferred quantities: a change in mass shifts the frequency scale, while changes in rotation, charge, or other continuous parameters can imitate a deformation of the potential barrier. It is therefore more informative to compare a model with the best-fitting member of the competing family than to compare curves at arbitrarily fixed parameter values. This can be expressed through the matched residual
\begin{equation}
\Delta\Gamma_{\ell m}^{AB}(\omega)
=\Gamma_{\ell m}^{(A)}(\omega;\boldsymbol{\lambda}_{A})
-\Gamma_{\ell m}^{(B)}(\omega;\widehat{\boldsymbol{\lambda}}_{B}),
\end{equation}
where $\boldsymbol{\lambda}_{A}$ and $\boldsymbol{\lambda}_{B}$ denote the intrinsic parameters of the two models and $\widehat{\boldsymbol{\lambda}}_{B}$ is chosen to give the best fit to model $A$ over the relevant frequency interval. An absolute residual is preferable to a relative one wherever $\Gamma_{\ell m}$ is close to zero, because an arbitrarily large percentage difference there may correspond to a negligible transmitted flux.

Parametrized geometries and parametrized effective potentials offer a systematic way to identify such degeneracies before testing particular theories. Analytic expansions for general parametrized spherically symmetric black holes show that, within their domain of validity, the grey-body factors are controlled predominantly by a few combinations of the deformation coefficients \cite{Konoplya:2020hyk}. In particular, the displacement of the horizon radius can have a larger effect than higher-order coefficients describing fine details of the near-horizon geometry \cite{Dubinsky:2024nzo}. A complementary parametrization directly at the level of the perturbation potential similarly shows how corrections with different radial powers and perturbative orders enter the transmission probability \cite{Antoniou:2026jvh}. Consequently, several theory parameters may generate almost parallel changes in the grey-body spectrum and cannot be constrained independently. The useful observational question is then which combinations of parametrized deformations are measurable, and only afterward which gravity theories map into the allowed region of that parameter space.

These considerations also determine how graphical comparisons should be presented. Grey-body factors for competing models should be plotted against the same dimensionless frequency after their common parameters have been matched or fitted. A companion panel should show the absolute matched residual $\Delta\Gamma_{\ell m}^{AB}$, while the frequency range that contributes appreciably to the detector response or to the thermal emission should be indicated. Several angular modes, field spins, or polarizations should be shown whenever available. Such a presentation separates genuine changes in the shape of the transition region from a simple horizontal rescaling, and it avoids visually magnifying relative differences in a low-frequency region where the transmission probability and observable flux are both negligible.

\begin{figure}[t]
\centering
\includegraphics[width=\linewidth]{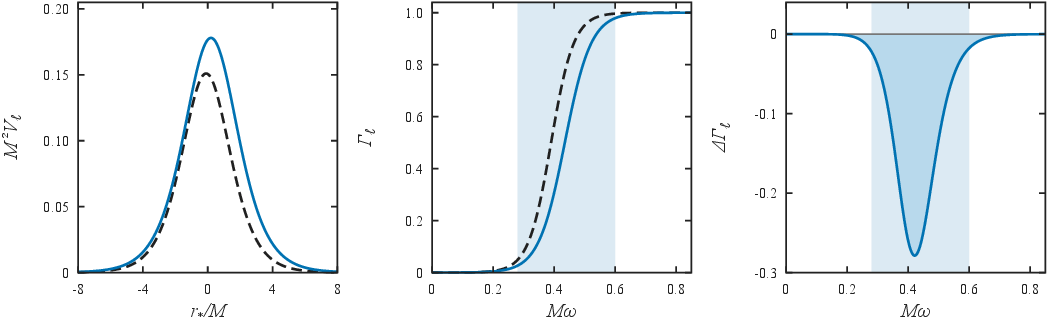}
\caption{Schematic comparison of scattering by a Schwarzschild black hole (dashed line) and an alternative black-hole model (solid line) for the same perturbation channel and multipole number. The left panel shows the effective potentials normalized by a common asymptotic mass $M$. The middle panel shows the corresponding grey-body factors. The shaded band indicates the intermediate-frequency transition region, where the curves generally carry the most model-dependent information. The right panel shows the signed additive difference $\Delta\Gamma_{\ell}=\Gamma_{\ell}^{\rm alt}-\Gamma_{\ell}^{\rm Schw}$.}
\label{fig:model_discrimination}
\end{figure}

Figure~\ref{fig:model_discrimination} illustrates this comparison schematically for one fixed perturbation channel and multipole number. A deformation of the effective barrier changes the location and width of the transition from almost complete reflection to almost complete transmission. The additive, rather than relative, difference between the grey-body factors is concentrated in this transition region and tends to zero at both ends of the plotted interval. Representative numerical results for axial gravitational perturbations, with specified multipoles and deformation parameters, are shown in Fig.~3 of Ref.~\cite{Konoplya:2023ppx} for regular scalar-hairy black holes and in Fig.~6 of Ref.~\cite{Antoniou:2025bvg} for scalarized black holes in shift-symmetric Horndeski gravity. Comparisons made using a geometric length scale, such as the horizon radius, can mix genuine changes in the scattering properties with a change in the overall scale. Thus, for models intended to describe the same physical black hole, the frequencies should therefore be rescaled using a common asymptotic mass, and any other shared physical charges should likewise be matched.

For gravitational-wave ringdown, the measured quantity is the waveform rather than an isolated grey-body factor. As discussed in Sec.~\ref{sec:ringdown_gbf}, the complex reflection amplitude is multiplied by source-dependent excitation terms, and these terms can be partially degenerate with changes in the remnant scattering response. The appropriate test is therefore a joint inference of the mass, rotation parameter, scattering response, and source parameters, followed by a detector-noise-weighted comparison of the resulting waveforms. Information from the inspiral or from standard quasinormal-mode spectroscopy can supply independent constraints on the remnant parameters. The \texttt{GreyRing} construction provides an example of this strategy by comparing the mass and angular momentum inferred from the grey-body response with those obtained from other portions or representations of the signal \cite{Oshita:2023cjz,Okabayashi:2024qbz,Rosato:2026apq}. Higher multipoles and the phase of the complex reflection amplitude can break degeneracies that remain in the dominant-mode spectral amplitude alone.

The corresponding degeneracies are different for Hawking radiation and absorption measurements. A Hawking spectrum contains the product of the grey-body factor and the thermal occupation factor, summed over all emitted species, so a modification of the transmission probability can be partly compensated by a change in the Hawking temperature, particle content, or decay and propagation of the emitted particles. An integrated luminosity loses still more spectral information. Discrimination therefore requires the frequency dependence and, where possible, several particle species or angular channels to be analyzed together (see Sec.~\ref{sec:Hawking}). Likewise, a total absorption cross section sums over partial waves and can hide compensating changes among them. Polarization-resolved or mode-resolved information is more diagnostic than a single total cross section.

Grey-body factors should therefore be treated as a family of frequency- and channel-dependent observables, not as a single label that uniquely specifies a compact object. A credible discrimination analysis should identify the frequency band and channels carrying the signal, fit the continuous parameters of every competing model, marginalize over source and propagation uncertainties, and quantify the remaining difference at the level of the measured spectrum or waveform. Combining several multipoles or particle species with independent measurements of the mass, angular momentum, shadow, or quasinormal spectrum is considerably more powerful than any one curve in isolation. In this role, grey-body factors provide a complementary consistency test of the strong-field geometry and perturbation dynamics, while the remaining degeneracies state precisely which aspects of the underlying theory cannot be inferred from scattering data alone.

\subsection{Wormhole transmission and superradiance}

Grey-body factors for traversable wormholes can be introduced in close analogy with the black-hole case once the scattering problem is formulated in terms of the tortoise coordinate. In this coordinate, the radial domain extends over $r_* \in (-\infty,+\infty)$, where the two limits correspond to two asymptotically flat regions belonging to different universes. As a result, the wave equation takes the same Schröinger-like form as for black holes, and the difference between the two systems is encoded in the effective potential and in the physical interpretation of the boundaries.

From the viewpoint of the scattering problem, a wormhole is therefore equivalent to a black hole written in the tortoise coordinate, with the event horizon replaced by another asymptotic region \cite{Konoplya:2005et}. Consequently, the boundary conditions used to define transmission and reflection coefficients have the same mathematical form as in black-hole scattering \cite{Konoplya:2010kv},
\begin{equation}
\begin{array}{rclcl}
\Psi &\to& e^{-i(\omega-\Phi_+) r_*}+R_{\ell m}e^{+i(\omega-\Phi_+) r_*}, &\quad& r_*\to+\infty, \\
\Psi &\to& T_{\ell m}e^{-i(\omega-\Phi_-)r_*}, &\quad& r_*\to-\infty.
\end{array}
\end{equation}

The boundary conditions are therefore formally similar to those used for black holes, but their interpretation is different. For a black hole, the transmitted wave is absorbed by the event horizon. For a traversable wormhole, there is no horizon in this scattering problem: the transmitted wave emerges in the other asymptotic region. Thus the grey-body factor of a wormhole measures the probability of passage through the throat, or transmission between the two exterior regions, rather than absorption by a horizon.

The quantities \(\Phi_+\) and \(\Phi_-\) denote the asymptotic frequency shifts in the two regions. They may be different, for example, if the Coulomb-potential asymptotics differ between the two regions, or if the two asymptotic regions rotate with respect to each other. With the convention used above, flux conservation gives (cf. Eq.~\ref{eq:greybodydS})
\begin{equation}
\Gamma_{\ell m}(\omega)\equiv1-|R_{\ell m}|^2 = \frac{\omega-\Phi_-}{\omega-\Phi_+}|T_{\ell m}|^2 .
\end{equation}
The wormhole analogue of superradiant amplification occurs when the frequency lies between $\Phi_-$ and $\Phi_+$. For a symmetric wormhole, however, the two asymptotic regions are equivalent and $\Phi_+=\Phi_-$, so the ratio in the flux-conservation relation is positive and no superradiant amplification occurs, in agreement with the analysis of symmetric wormholes in Ref.~\cite{Konoplya:2010kv}.

Depending on the shape of the effective potential, which may exhibit a double-barrier structure, the transmission spectrum can display features such as resonances and quasi-bound states associated with wave trapping near the throat. Such effects were analyzed in \cite{Konoplya:2025hgp}, where it was shown that wormhole geometries can lead to characteristic oscillatory patterns in the transmission coefficients. Thus, although the formal definition of grey-body factors remains unchanged, their physical interpretation and phenomenology differ significantly from the black-hole case. For ultracompact horizonless objects, the same scattering response links long-lived modes and their excitation factors to echoes and grey-body factors \cite{Rosato:2025byu,Rosato:2025lxb}.

\subsection{Grey-body factors in various theories of gravity}

Grey-body factors for black holes in a wide variety of gravitational theories and matter configurations have been investigated in numerous studies, which are summarized in Table~\ref{tabl:grey-body}. We note, however, that the literature on the subject is considerably broader than the collection presented there. Many works, omitted from Table~\ref{tabl:grey-body}, are limited either to analytical low-frequency approximations or to estimates based solely on the Visser bound (see Sec.~\ref{sec:Visser}), without providing a detailed characterization of the transmission spectrum over a wider frequency range. For inclusion in the table, we adopted a more restrictive criterion: the analysis must contain either an accurate numerical computation of the grey-body factors or, at the very least, a WKB-based estimate of the transmission coefficients. This requirement ensures that the cited studies provide quantitative information on the frequency dependence of the grey-body factors beyond their asymptotic low-energy behavior.

The emphasis on obtaining sufficiently accurate grey-body factors is motivated by two important considerations. First, recent studies have demonstrated a close connection between grey-body factors and quasinormal modes, making it possible to extract information about the ringdown spectrum from scattering characteristics (see Sec.~\ref{sec:correspondence}). Since this correspondence relies on relatively subtle features of the transmission spectrum, high-precision calculations of grey-body factors are essential for meaningful comparisons and applications.

Second, accurate grey-body factors are crucial for reliable estimates of Hawking radiation (see Sec.~\ref{sec:Hawking}). Although the Hawking temperature is often the primary quantity controlling the overall evaporation rate, the transmission probabilities may have a comparable impact on the resulting emission spectrum \cite{Konoplya:2019ppy}. In some situations, modifications of the grey-body factors can even compensate for or outweigh the effect of changes in temperature. This issue becomes particularly important in cases where the black hole possesses the same Hawking temperature as the Schwarzschild solution, so that any difference in the Hawking radiation spectrum originates entirely from the grey-body factors themselves \cite{Konoplya:2023bpf}. Consequently, the precise determination of the transmission coefficients is indispensable for distinguishing between different black-hole geometries through their evaporation signatures.

\begin{table*}
%\centering
\begin{tabular}{p{0.48\columnwidth}|p{0.13\columnwidth}|p{0.13\columnwidth}|p{0.14\columnwidth}}
%Theory / framework
%\multicolumn{4}{c}{Representative papers on grey-body factors} \\
 \hline
\multicolumn{1}{r|}{Approach:} & semi-analytic & WKB & numerical \\

\hline

Schwarzschild black hole & \cite{Page:1976df,Sanchez:1976} & \cite{Sanchez:1978} & \cite{Page:1976df}\\

Kerr black hole & \cite{Macedo:2013afa} & & \cite{Page:1976ki,Dolan:2008kf,Macedo:2013afa,Leite:2017zyb,Leite:2018mon} \\

Reissner--Nordström black hole & \nocite{Malik:2026jzl} \cite{Benone:2014qaa,Benone:2015bst} & \cite{Malik:2026jzl} & \cite{Nakamura:1976nc,Page:1977um,Matzner:1976kj,Crispino:2009ki,Crispino:2009zza,Benone:2014qaa,Benone:2015bst} \\

Kerr--Newman black hole & \cite{Maulik:2025hax,Benone:2018rtj,Benone:2019all,Sharif:2020hyz} & & \cite{Leite:2017hkm,Benone:2018rtj,Kokkotas:2010zd}\\

Parametrized black holes & \cite{Dubinsky:2024nzo} & \cite{Dubinsky:2024nzo,Antoniou:2026jvh} & \cite{Antoniou:2026jvh} \\

Black holes immersed in a galactic halo & &\cite{Konoplya:2021ube,Bolokhov:2026eqf,Skvortsova:2026unq,Lutfuoglu:2025mqa,Lutfuoglu:2025kqp} & \\

Euler--Heisenberg black holes & & \cite{Malik:2025erb,Silva:2026qlr} & \cite{Tan:2026itp}\\

Stringy black holes & \cite{Xavier:2021sje} &\cite{Dubinsky:2024vbn} & \cite{Xavier:2021sje} \\

\hline

Higher-dimensional black holes & \cite{Kanti:2002nr,Harris:2003eg,Cardoso:2005vb,Cardoso:2005mh,Harmark:2007jy,Cvetic:2009jn} & &\cite{Harris:2003eg,Harris:2005jx,Jung:2005mr,Casals:2008pq}\\

Higher-dimensional black holes projected on the brane & \cite{Kanti:2002nr,Kanti:2002ge,Ida:2002ez} & \cite{Dubinsky:2025ypj} & \cite{Duffy:2005ns,Casals:2005sa,Jung:2005mr,Casals:2006xp}\\

Einstein--Gauss--Bonnet / Lovelock gravity & \cite{Zhang:2017yfu} &\cite{Konoplya:2010vz,Konoplya:2020cbv} & \cite{Grain:2005my,Zhang:2020qam,Tan:2026vif} \\

Higher-dimensional string-corrected black holes & \cite{Moura:2024vhz} & & \cite{Grain:2005my}\\

\hline

Gravity with higher-derivative corrections & & \cite{Konoplya:2019ppy} & \cite{Moura:2024vhz}\\

$f(R)$ gravity / cloud of strings backgrounds & \cite{Ovgun:2018gwt} & & \\

Scalar-tensor, Horndeski and beyond-Horndeski theories & & \cite{Konoplya:2023ppx,Bolokhov:2026uol,Konoplya:2026cus,Lutfuoglu:2026gey} & \cite{Antoniou:2025bvg}\\

Einstein--scalar--Gauss--Bonnet / scalarized black holes & &\cite{Konoplya:2019hml,Paul:2023eep} & \cite{Antoniou:2025bvg}\\

Black holes in Proca gravity & &\cite{Lutfuoglu:2026pgn,Lutfuoglu:2025ldc} & \\

\hline

Regular (quantum corrected) black holes & & \cite{Malik:2025dxn,Dubinsky:2025wns,Dubinsky:2025nxv,Lutfuoglu:2026rqe} & \cite{3164933,Konoplya:2023ahd,Pathrikar:2025gzu,Calza:2024fzo,Calza:2024xdh,Calza:2025mwn,Hui:2026tsn,Skvortsova:2026mwz} \\

Regular black holes in Asymptotically Safe Gravity & &\cite{Konoplya:2023bpf,Stashko:2024wuq,Lutfuoglu:2025ohb,Malik:2025qnr,Bolokhov:2025lnt,Lutfuoglu:2025blw} &\\

Regular black holes in scalar-tensor conformal gravity & & \cite{Li:2021epb} & \\

Regular black holes in Einstein-Dirac-Born-Infeld theory & &\cite{Bolokhov:2026kqu} & \\

Regular black holes in quasi-topological gravity & &\cite{Arbelaez:2026eaz,Arbelaez:2026cnd,Dubinsky:2026wcv} & \cite{Konoplya:2025uta} \\

\hline

Black holes in Lorentz-violating gravity & &\cite{Guo:2023nkd} & \\

dRGT massive gravity / massive black strings & & \cite{Boonserm:2023oyt} & \cite{Dong:2015qpa} \\

Hořava--Lifshitz gravity & \cite{Ding:2010kr} & & \\

Rastall gravity & &\cite{Karmakar:2023cwg,Gogoi:2023lvw} & \\

Symmergent gravity& &\cite{Gogoi:2023fow,Gogoi:2024epx} & \\

T-duality and self-energy & &\cite{3164933} & \cite{3164933}\\

Effective Field Theories & &\cite{Konoplya:2023ppx} & \\

Charged accelerating black holes & &\cite{Moura:2026ken} &\\

Holonomy corrected black holes & & \cite{Lutfuoglu:2026vcr} & \cite{Lutfuoglu:2026vcr}\\

\hline

Wormholes & &\cite{Konoplya:2010kv,Taylor:2014vsa,Malik:2024wvs,Bolokhov:2024otn} &\cite{Rosato:2025byu,Karmakar:2026tpa,Riley:2026avn}\\

Black holes evaporating into wormholes & & & \cite{Lutfuoglu:2026zxj}\\

\hline
\end{tabular}
\caption{Representative papers on grey-body factors of black holes and wormholes.}\label{tabl:grey-body}
\end{table*}

Although grey-body factors are frequently viewed as corrections to the thermal spectrum whose role is subordinate to that of the Hawking temperature, this is not always the case. An illustrative example of the complex interplay between the Hawking temperature and grey-body factors is provided by non-Schwarzschild black holes in higher-derivative gravity \cite{Konoplya:2019ppy}. Although the Hawking temperature decreases monotonically as the coupling constant grows, the Hawking radiation intensity is initially enhanced rather than suppressed. This counterintuitive behavior is explained by a substantial increase in the transmission probabilities caused by the lowering of the effective potential barrier. Only for sufficiently large couplings does the decrease in temperature overcome the grey-body-factor enhancement and lead to suppression of the radiation spectrum. Consequently, grey-body factors cannot always be regarded as a secondary correction to the thermal spectrum and may play a decisive role in determining the evaporation rate of a black hole.

The importance of grey-body factors was further highlighted in studies of higher-dimensional Gauss--Bonnet black holes. It was shown that even relatively small higher-curvature corrections may increase the black-hole lifetime by several orders of magnitude \cite{Konoplya:2010vz}. This dramatic suppression of Hawking radiation originates from the combined action of two effects: the reduction of the Hawking temperature and the simultaneous decrease of the grey-body factors, particularly for tensor gravitational perturbations. As a consequence, estimates based solely on the thermal spectrum or on the approximation of unit grey-body factors may significantly overestimate the evaporation rate and underestimate the lifetime of black holes.

A related and particularly important question concerns the reliability of truncating the infinite tower of higher-curvature corrections expected in ultraviolet completions of gravity, such as string-inspired effective theories. While any finite-order theory represents only an approximation to the full resummed geometry, it was recently shown that the Hawking radiation spectrum of regular black holes constructed from an infinite series of quasi-topological curvature terms \cite{Bueno:2024dgm} converges surprisingly rapidly as successive orders are included \cite{Konoplya:2025uta}. Although quantities sensitive to the near-horizon region, such as grey-body factors and Hawking emission rates, require more terms for quantitative accuracy than classical observables associated with the peak of the effective potential \cite{Konoplya:2024hfg}, only a modest number of higher-curvature corrections is sufficient to reproduce the predictions of the full infinite-series solution with good precision. This result provides strong evidence that finite-order higher-curvature theories can capture the essential evaporation properties of their ultraviolet-completed counterparts.

A different perspective on the role of grey-body factors emerges in the context of higher-dimensional black holes. Extensive studies by P. Kanti and collaborators have shown that, although accurate grey-body factors are indispensable for determining the detailed emission spectra and the partition of radiation between the brane and the bulk, the overall evaporation process is strongly influenced by the rapidly increasing Hawking temperature and the large multiplicities of higher-dimensional particle modes. As the number of extra dimensions grows, these effects can enhance the total emission rates by orders of magnitude, significantly shortening the black-hole lifetime. Therefore, while grey-body factors govern the transmission of individual modes through the effective potential barrier, the total Hawking output may be controlled to a large extent by the available phase space and the multiplicity of radiative degrees of freedom \cite{Kanti:2004nr,Kanti:2005ja,Kanti:2008eq,Harris:2003eg,Kanti:2014vsa}.

\section{Grey-body factors and the emergence of gauge/gravity duality}\label{sec:duality}

In string theory, black branes arise as higher-dimensional generalizations of black holes, typically associated with extended objects (D-branes) on which open strings can end \cite{Akhmedov:2001un}. A crucial development was the realization that, in certain limits, the near-horizon geometry of such branes becomes asymptotically anti-de Sitter (AdS), while their low-energy dynamics is described by a quantum field theory living on the brane worldvolume. This observation led to the formulation of the AdS/CFT correspondence, which posits an equivalence between a gravitational theory in a higher-dimensional AdS spacetime and a conformal field theory (CFT) defined on its boundary. Within this framework, classical processes such as wave propagation and absorption by black branes acquire a dual interpretation in terms of correlation functions and excitations in the boundary theory \cite{Kovtun:2004de}.

Grey-body factors, understood as transmission coefficients for wave propagation in curved spacetime, gained particular prominence in the late 1990s when they appeared in a completely different context: the study of absorption by black branes in string theory. In this setting, one considers the propagation of a test field in a higher-dimensional gravitational background and computes the corresponding absorption probability by solving a radial wave equation of the form (\ref{eq:schroedinger_simple}) subject to the same purely ingoing boundary conditions at the horizon and a superposition of incoming and outgoing waves at infinity (\ref{eq:scattering}).

The early D-brane calculations compared two descriptions of the same low-energy absorption process. On the gravitational side, one solves the wave equation for scalar perturbations in a near-extremal black-hole or black-brane background and extracts the frequency-dependent grey-body factor by matching the near-horizon and asymptotic solutions. In the D-brane description, the same process is represented by transitions among excitations of a brane bound state, or equivalently of an effective string. The agreement obtained in Ref.~\cite{Maldacena:1996ix} was important because it reproduced not only the general existence of Hawking emission, but also the detailed frequency dependence of the grey-body spectrum. In this sense, grey-body factors provided one of the first quantitative tests that semiclassical radiation from a black hole can be described in quantum/string-theory terms.

This line of work was then developed in several directions. Four-dimensional grey-body factors were reproduced in an effective-string framework, where the absorption probabilities follow from finite-temperature dynamics of the string degrees of freedom \cite{Gubser:1996zp}. Scalar absorption by effective strings further clarified how the gravitational matching calculation is mirrored by worldvolume interaction vertices and thermal occupation numbers \cite{Klebanov:1997cx}. The comparison was also extended to broader families of charged and rotating black holes in string theory \cite{Cvetic:1997uw,Cvetic:1997xv}. In these backgrounds the grey-body factors depend on the charges, angular momenta, and effective left- and right-moving temperatures that characterize the quantum/string-theory description. The resulting expressions showed how rotation and charge modify the absorption spectrum, including the frequency dependence and the channels associated with energy and angular-momentum exchange.

The key observation made in \cite{Gubser:1997yh,Klebanov:1997kc} is that, for certain near-extremal black brane geometries, the same absorption cross section can be computed independently in a dual description. On the gravitational side, one solves the above wave equation in the curved background and extracts $\Gamma_\ell(\omega)$ using standard scattering techniques. On the field-theory side, the absorption process is described by the interaction of external perturbations with degrees of freedom living on the brane worldvolume. The corresponding absorption probability is proportional to the imaginary part of a retarded two-point function,
\begin{equation}
\sigma_{\text{abs}}(\omega) \propto \frac{1}{\omega} \,\mathrm{Im}\, G_R(\omega),
\end{equation}
where $G_R(\omega)$ is the retarded correlator of the operator coupled to the bulk field.

The remarkable result is that these two seemingly unrelated computations yield identical expressions for the absorption cross section in the low-frequency regime. In particular, for minimally coupled scalar fields, one finds that at leading order
\begin{equation}
\sigma_{\text{abs}}(\omega) \to A_H,
\end{equation}
where $A_H$ is the horizon area of the black brane. This universality, derived from classical gravity, is exactly reproduced by the worldvolume theory, providing a nontrivial consistency check between the two descriptions.

This agreement goes far beyond a coincidence. It indicates that the grey-body factor, obtained from a classical scattering problem in a curved background, encodes the same physical information as correlation functions in a quantum field theory without gravity. From the modern perspective, this relation is understood as a manifestation of gauge/gravity duality, where bulk propagation is mapped to boundary dynamics \cite{Aharony:1999ti,Gubser:1998bc,Witten:1998qj,Maldacena:1997re}. In this framework, the transmission probability $\Gamma_\ell(\omega)$ determines the spectral density of the dual theory, and the effective potential barrier in the bulk plays the role of an energy-dependent filter for excitations.

The same logic underlies the use of black-hole backgrounds in holographic superconductors. In the simplest construction, a charged scalar field coupled to a Maxwell field is placed in an asymptotically AdS black-hole spacetime. Below a critical temperature the scalar develops a nontrivial radial profile, which is interpreted in the boundary theory as the condensation of a charged operator and the spontaneous breaking of the global \(U(1)\) symmetry \cite{Hartnoll:2008vx,Herzog:2009xv}. Linear electromagnetic perturbations then obey a radial wave equation with an ingoing boundary condition at the horizon and a prescribed source at the AdS boundary. The optical conductivity of the dual system is extracted from the corresponding retarded Green function, so the calculation is again a connection problem for a bulk wave propagating through an effective potential.

From the grey-body-factor viewpoint, the condensate modifies the effective potential seen by the electromagnetic perturbation and therefore changes the amount of flux absorbed by the horizon. In the boundary theory this change appears as a frequency-dependent conductivity, including the formation of a gap and a redistribution of spectral weight in the superconducting phase. In the zero-temperature limit the conductivity is controlled by tunneling through the effective bulk barrier rather than by a simple thermal absorption law \cite{Horowitz:2009ij,Konoplya:2009hv}. Thus holographic superconductors provide a concrete example in which the grey-body-factor intuition is translated into a transport coefficient of a strongly coupled quantum system.

An important feature of these results is that they rely on the existence of a well-defined matching region in which analytic approximations are valid. In the low-energy limit, $\omega r_H \ll 1$, the wave equation can be solved in different regions and matched, leading to explicit expressions for $\Gamma_\ell(\omega)$ (see Sec.~\ref{sec:lowfrequencymatching}). This analytic control made it possible to demonstrate the precise agreement between the gravitational and field-theoretical calculations, thereby providing one of the earliest quantitative indications of the correspondence between the two descriptions.
In this way, grey-body factors probe the response of the geometry to external perturbations in a way that is sensitive to both near-horizon physics and asymptotic structure. Their appearance in the early tests of gauge/gravity duality underscores their deeper significance: they serve as a bridge between classical scattering theory and quantum dynamics, linking wave propagation in curved spacetime to the behavior of strongly coupled quantum systems.

\section{Summary and outlook}\label{sec:summary}

Grey-body factors constitute one of the central ingredients of black-hole physics, linking the near-horizon generation of radiation with observations performed at asymptotic infinity. They determine the modification of Hawking radiation by spacetime curvature, govern absorption and scattering processes, and encode important information about the effective potential surrounding compact objects. Although originally introduced in the context of black-hole thermodynamics, grey-body factors have evolved into a versatile tool applicable to a broad spectrum of problems ranging from black-hole evaporation and wave scattering to holography and, more recently, black-hole spectroscopy \cite{Rosato:2026apq,Rosato:2024arw,Okabayashi:2024qbz}.

In this review we have summarized the theoretical foundations of grey-body factors, beginning with the formulation of the scattering problem under various asymptotic boundary conditions. We discussed the relation between transmission coefficients, absorption cross sections, and Hawking radiation, and reviewed the principal analytic, semi-analytic, and numerical techniques used for their calculation. Particular attention was devoted to the WKB approach, low- and high-frequency approximations, numerical integration methods, and modern spectral techniques. We also surveyed the rapidly growing literature devoted to grey-body factors in black holes, wormholes, black branes, and compact objects arising in modified theories of gravity.

Over the past decades, substantial progress has been made in understanding the role of grey-body factors in black-hole physics. Accurate calculations are now available for a wide variety of backgrounds, including higher-dimensional, rotating, charged, regular, and quantum-corrected black holes. At the same time, several important challenges remain. Reliable calculations in geometries with multiple potential barriers, rapidly varying effective potentials, or nontrivial asymptotic structures continue to require further methodological development. Likewise, obtaining high-precision results for rotating black holes, especially beyond the separable Kerr family, remains a demanding task.

Using grey-body factors as observables introduces a related set of difficulties. Although the intrinsic grey-body factors are generally stable under weak environmental deformations over the relevant frequency ranges, matter within the scattering region can modify the effective potential, while absorption or reprocessing during propagation can alter the signal received by a distant observer. Furthermore, the inverse problem is not unique: different effective potentials, parameter choices, and source or propagation effects can produce nearly degenerate spectra over a finite frequency band. Reliable inference therefore requires consistent modeling of the environment and source, comparisons across the full observable band, and, where possible, several multipoles or particle species together with independent constraints on the compact object's parameters.

One of the most intriguing recent developments is the discovery of a correspondence between grey-body factors and quasinormal modes, which we discussed in Sec.~\ref{sec:correspondence}. This connection suggests that scattering data may provide information about the ringdown spectrum and, conversely, that spectral properties of black holes can be inferred from transmission probabilities. Current investigations indicate that the correspondence is remarkably robust in a broad class of situations, although its precise domain of validity is still being established. Clarifying the underlying physical mechanism and determining the conditions under which the correspondence breaks down represent important directions for future research \cite{Pedrotti:2025idg}.

Another rapidly developing area concerns the role of grey-body factors in Hawking radiation and black-hole evaporation. While the Hawking temperature often provides the dominant contribution to the emission rate, recent studies have demonstrated that grey-body factors may substantially alter the resulting spectra and, in some cases, become the primary source of observable differences between competing black-hole models. This makes the accurate determination of transmission probabilities increasingly important for phenomenological studies of primordial black holes, regular and higher-dimensional black holes, and quantum-gravity-inspired compact objects \cite{Kanti:2004nr,Spina:2025wxb}.

Grey-body factors have also acquired a prominent role in gauge/gravity duality, where absorption probabilities and scattering amplitudes provide a direct bridge between gravitational dynamics in the bulk and correlation functions in the dual field theory. Further progress in this direction may deepen our understanding of black-hole microphysics and the quantum aspects of gravity \cite{Klebanov:1997kc,Gubser:1997cm,Maldacena:1997re,
Aharony:1999ti}.

Looking forward, the continued development of analytical methods, numerical techniques, and observational probes will likely further broaden the role of grey-body factors in gravitational physics. Their unique ability to connect black-hole thermodynamics, scattering theory, quasinormal modes, and holography suggests that they occupy a special position among black-hole observables. As increasingly sophisticated models of compact objects are developed and future observations provide tighter constraints on strong-gravity phenomena, grey-body factors are expected to remain an indispensable tool for exploring both classical and quantum aspects of gravitation.

\ack{AZ was partially supported by Conselho Nacional de Desenvolvimento Científico e Tecnológico (CNPq).
The authors acknowledge the assistance of OpenAI in reviewing language during the preparation of this manuscript. The scientific content, interpretations, and final text are the responsibility of the authors.}

\bibliographystyle{iopart-num}
\setlength{\bibsep}{2pt}
\bibliography{bibliography}
\end{document}